\documentclass{JFM-FLM_Au}
\usepackage{bm}

\lefttitle{A. Xu, H.-L. Wu, B.-R. Xu and H.-D. Xi}
\righttitle{Journal of Fluid Mechanics}

\title{Learning to traverse convective flows at moderate to high Rayleigh numbers}

\author{Ao Xu\aff{1,2}, Hua-Lin Wu\aff{1}, Ben-Rui Xu\aff{1} \and Heng-Dong Xi\aff{1,2}}

\affiliation{\aff{1}Institute of Extreme Mechanics, School of Aeronautics, Northwestern Polytechnical University, Xi'an 710072, PR China
\aff{2}National Key Laboratory of Aircraft Configuration Design, Key Laboratory for Extreme Mechanics of Aircraft of Ministry of Industry and Information Technology, Xi'an 710072, PR China}

\corresau{Heng-Dong Xi, \email{hengdongxi@nwpu.edu.cn}}

\begin{document}
\maketitle

\begin{abstract}
We study the navigation of a self-propelled inertial particle in two-dimensional Rayleigh--B\'enard convection at Prandtl number \(Pr=0.71\) and cell aspect ratio \(\Gamma=4\) for Rayleigh numbers \(Ra\) ranging from \(10^7\) to \(10^{11}\). A reinforcement-learning (RL) controller selects the propulsive acceleration, subject to an upper bound \(\mathcal{A}_{\max}\), to achieve a prescribed horizontal displacement. We find that the success rate increases abruptly with \(\mathcal{A}_{\max}\) at moderate \(Ra\), whereas at higher \(Ra\) the transition becomes more gradual and shifts to larger \(\mathcal{A}_{\max}\). Moreover, although the completion time increases with \(Ra\), the propulsion energy required for successful traversal decreases. Proper orthogonal decomposition indicates that these performance differences are associated with reorganisation of the carrier flow. At moderate \(Ra\), the dominant large-scale circulation partitions the domain through persistent transport barriers, requiring a finite thrust surplus to cross them; at higher \(Ra\), energy is distributed across many modes, the barriers fragment and transient plume-assisted pathways emerge. Compared with a constant-heading baseline, the learned policy aligns with local currents and consumes significantly less energy. Lagrangian coherent structure analysis further suggests that the RL agent tends to cross repelling barriers and surf along attracting pathways. Finally, by mapping these behaviours onto the local Eulerian flow topology using Voronoi tessellation and the \(Q\)-criterion, we distil an interpretable, physics-based heuristic strategy that retains robust navigability. These results connect turbulent-flow organisation with autonomous navigation under bounded actuation.
\footnote{
This article may be downloaded for personal use only.
Any other use requires prior permission of the author and Cambridge University Press.
This article appeared in Xu \emph{et al.}, J. Fluid Mech. \textbf{1039}, A18 (2026) and may be found at \url{https://doi.org/10.1017/jfm.2026.11802}.
}
\end{abstract}


\section{Introduction}
\label{sec:Introduction}

Self-propelled particles span length scales from micrometres to metres and harvest energy from their surroundings to generate motion \citep{Bechinger2016RMP}. 
Natural examples include molecular motors, cells, bacteria, fish and birds, which are powered by chemical or biological energy \citep{Gustavsson2016PRL, Sengupta2017Nature, Qiu2022ActaMS}. Artificial counterparts include self-propelled microparticles and macroscopic robots driven by thermal or electrical energy \citep{Cichos2020NMI}. Trajectory planning is crucial in many settings, including microorganisms navigating turbulent flows \citep{Monthiller2022PRL, Qiu2022PRF}, cells migrating along chemical gradients \citep{Sengupta2017Nature} and birds undertaking long-distance flights by thermal soaring \citep{Weimerskirch2016Science}. 
These navigation problems can be classified according to the environmental information used to guide motion. 
Short-range navigation may rely on sensing hydrodynamic disturbances \citep{Qiu2022JFM, Borra2022PRF, Zhu2022JFM, Xu2022PoF}, medium-range navigation on chemotaxis or infotaxis \citep{Vergassola2007Nature, Loisy2022PRSA, Rigolli2022eLife} and long-range navigation on coupled horizontal and vertical dispersion \citep{Colabrese2017PRL, Gustavsson2017EPJE, Biferale2019Chaos, Gunnarson2021NC, Xu2023PRF}. 
A substantial body of work on Lagrangian navigation builds on Zermelo's problem of determining optimal routes in unsteady currents using level-set methods, graph search and stochastic control. 
Because models and observations are often imperfect, recent studies have incorporated estimation and adaptation into the planning process through learning-based flow prediction, belief-space planning and reinforcement-learning (RL) control \citep{simons2004many,piro2025many,heinonen2025exploring}.
Compared with traditional model predictive control, which requires computationally expensive forward prediction and is highly sensitive to the exponential divergence of trajectories in spatiotemporally chaotic flows \citep{Krishna2022PRSA,krishna2023finite}, model-free RL offers particular advantages. It can learn an instantaneous reactive policy from local sensing, thereby reducing the need for real-time flow forecasting. Furthermore, understanding how well such learned policies generalise across flow regimes is important for practical deployment \citep{Jiao2025Gradients,hang2026flow}.

Investigations of self-propelled particles in natural systems have also inspired the design of artificial swimmers and autonomous vehicles for complex environments. For example, \cite{MuinosLandin2021SR} studied a self-thermophoretic microswimmer in real time and found that its learning efficiency decreased under strong Brownian noise. Related studies have also examined the control of autonomous vehicles in aerial and marine environments. \cite{Reddy2016PNAS} simulated a glider ascending in a spiral pattern reminiscent of soaring birds \citep{Akos2008PNAS}; this concept was later demonstrated in field experiments, in which a two-metre glider navigated autonomously using atmospheric thermals \citep{Reddy2018Nature}. \cite{Bellemare2020Nature} deployed balloons in the stratosphere that used wind speed, altitude and direction for station keeping and travel, enabling long-duration monitoring of air quality and environmental variables. \cite{Masmitja2023SR} developed an autonomous surface vehicle capable of locating and tracking moving underwater targets, and validated the method through sea trials involving several autonomous platforms, including the Sparus II autonomous underwater vehicle.

In both atmospheric and oceanic environments, convective currents play a key role in determining navigation paths that minimise time or energy \citep{Biferale2019Chaos, Laurent2021PNAS}. These flows are characterised by buoyancy-driven turbulence spanning a wide range of scales, presenting both opportunities and challenges for navigation. Intermittent updrafts and cross-stream currents can reduce propulsion requirements, but three main features of convective turbulence complicate control: multi-scale variability, which demands continuous sensing and adaptation across a broad range of spatial and temporal scales; rapid decorrelation, which shortens predictability horizons; and spatial intermittency, which generates localised updrafts and preferred pathways. To analyse such convective flows, we use Rayleigh--B\'enard (RB) convection as an idealised canonical model for buoyancy-driven convection relevant to atmospheric and oceanic flows \citep{Lohse2010ARFM, Chilla2012EPJE, Pandey2018NC, Wang2020SA, Xia2023NSR, Lohse2023PT, Lohse2024RMP, Shishkina2024PRL, Xia2025ActaMS}.

The primary control parameters of the RB system are the Rayleigh number (\(Ra\), which measures buoyancy relative to viscous and thermal diffusion effects), the Prandtl number (\(Pr\), which characterises the thermophysical properties of the fluid) and the cell aspect ratio (\(\Gamma\), which defines the geometry of the convection domain). At moderate \(Ra\), the flow organises into a roll-like large-scale circulation (LSC) that persists over many convective times. Boundaries between neighbouring cells, where the mean circulation reverses, are relatively well defined. Because plumes and roll edges remain coherent in space and time, transport is channelled through a few preferred pathways across the domain. As \(Ra\) increases, plume production intensifies, coherent rolls weaken and the interior develops richer small-scale structure \citep{Zhu2018PRL, Samuel2022PoF}. Temporal correlations of velocity and temperature then decay more rapidly on convective time scales, and transport pathways become less uniform because plume activity is increasingly intermittent. These regime changes matter for guidance and control because they affect the navigability of the flow under a fixed actuation limit. Specifically, flows dominated by LSC exhibit distinct pathways and barriers that influence reachability and the actuation required to move between basins, whereas flows at higher \(Ra\) offer more diffuse routes whose availability varies on short time scales. This motivates a closer investigation of how a self-propelled particle adapts its propulsive acceleration as turbulence intensity varies.

In thermal soaring and energy-harvesting flight, heuristic and optimal controllers exploit updrafts to extend flight range. More recently, deep reinforcement learning has been applied to flow control and to navigation in unsteady or turbulent environments, demonstrating flow-aware behaviour under a fixed actuation limit. However, most existing studies consider a single flow intensity and assess performance primarily in terms of time to target or qualitative success. It remains unclear how turbulence intensity determines reachability thresholds, shapes the trade-off between energy and time and influences the performance of learned guidance policies. In this work, we address this question by studying the navigation of self-propelled particles in convective environments at moderate to high Rayleigh numbers. Because classical optimal-navigation methods can be sensitive to small disturbances in chaotic flows, we adopt a data-driven approach based on an RL algorithm for trajectory optimisation \citep{Mehta2019PR, Brunton2020ARFM}. In recent years, RL has been increasingly used in the active-matter community, from early demonstrations of RL-driven active particles in simple environments \citep{schneider2019optimal} to more complex navigation and obstacle-avoidance tasks \citep{nasiri2022reinforcement}, as well as the discovery of energy-efficient swimming gaits \citep{ZhuKangTongEtAl2025}. The rest of this paper is organised as follows. In \S~\ref{sec:NumericalMethod}, we present the numerical methodology and the RL framework. In \S~\ref{sec:ResultsDiscussion}, we evaluate navigation performance in terms of reachability, completion time and energy expenditure across Rayleigh numbers and actuation limits. We then interpret these trends using proper orthogonal decomposition (POD) and Lagrangian coherent structures (LCS) to examine how the agent balances energy trade-offs and traverses transport barriers, and map the learned policy onto the local Eulerian flow topology to distil an interpretable heuristic strategy. Finally, in \S~\ref{sec:Conclusion}, we summarise the main findings of the present work.

\section{Numerical method}
\label{sec:NumericalMethod}

\subsection{Direct numerical simulation of thermal turbulence}

We consider incompressible thermal convection under the Boussinesq approximation, in which temperature acts as an active scalar and enters the momentum equation through buoyancy. All transport coefficients are taken to be constant. The governing equations are

\begin{align}
\nabla \cdot \bm{u}_f &= 0, \label{eq:conti}\\[3pt]
\frac{\partial \bm{u}_f}{\partial t} + \bm{u}_f \cdot \nabla \bm{u}_f &= -\frac{1}{\rho_0} \nabla P + \nu \nabla^2 \bm{u}_f + g \beta (T - T_0)\,\hat{\bm{y}}, \label{eq:mom}\\[3pt]
\frac{\partial T}{\partial t} + \bm{u}_f \cdot \nabla T &= \alpha \nabla^2 T, \label{eq:temp}
\end{align}
where $\bm{u}_f$ is the fluid velocity and $P$ and $T$ are the fluid pressure and temperature, respectively. The parameters $\beta$, $\nu$ and $\alpha$ denote the thermal expansion coefficient, kinematic viscosity and thermal diffusivity, respectively. The subscript `$0$' indicates a reference value, and $g$ is the magnitude of the gravitational acceleration. 
The unit vector $\hat{\bm{y}}$ points opposite to the direction of gravity.

We introduce the following dimensionless variables:

\begin{equation}
    \begin{split}
& \bm{x}^* = \frac{\bm{x}}{H}, \quad
t^* = \frac{t}{\sqrt{H/(g\beta \Delta_T)}}, \quad
\bm{u}_f^* = \frac{\bm{u}_f}{\sqrt{g\beta \Delta_T H}}, \\
& P^* = \frac{P}{\rho_0 g \beta \Delta_T H}, \quad
T^* = \frac{T - T_0}{\Delta_T},
    \end{split}
\end{equation}
With these definitions, \eqref{eq:conti}--\eqref{eq:temp} become, in dimensionless form,

\begin{align}
\nabla \cdot \bm{u}_f^* &= 0, \label{eq:contid}\\[3pt]
\frac{\partial \bm{u}_f^*}{\partial t^*} + \bm{u}_f^* \cdot \nabla \bm{u}_f^* &= -\nabla P^* + \sqrt{\frac{Pr}{Ra}}\,\nabla^2 \bm{u}_f^* + T^* \hat{\bm{y}}, \label{eq:momd}\\[3pt]
\frac{\partial T^*}{\partial t^*} + \bm{u}_f^* \cdot \nabla T^* &= \sqrt{\frac{1}{Pr\,Ra}}\,\nabla^2 T^*, \label{eq:tempd}
\end{align}
where \(H\) is the cell height and \(\Delta_T\) is the temperature difference between the heated bottom wall and the cooled top wall. The two dimensionless control parameters appearing in the governing equations are the Rayleigh number, \(Ra\), and the Prandtl number, \(Pr\), defined as

\begin{equation}
Ra = \frac{g \beta \Delta_T H^3}{\nu \alpha}, \qquad
Pr = \frac{\nu}{\alpha}. \label{eq:RaPr}
\end{equation}

We perform the direct numerical simulations using the spectral element method implemented in the open-source Nek5000 solver (version~v19.0). This method is suitable for the present problem because its high-order spatial accuracy and low numerical dissipation allow us to resolve a broad range of turbulent scales, capture fine intermittent structures and maintain numerical stability at high Rayleigh numbers (\(Ra \ge 10^{10}\)).
In Nek5000, the effective resolution in each spatial direction is determined by the product of the number of spectral elements and the polynomial order. 
Here, we use polynomial order $N = 9$ for the spectral-element discretisation. 
The simulations span the range $10^{7} \le Ra \le 10^{11}$ at fixed Prandtl number $Pr = 0.71$. 
To ensure adequate spatial and temporal resolution, we perform an \emph{a posteriori} verification by examining the number of grid points within the thermal boundary layer, the maximum wall-normal grid spacing $(\Delta_{g})_{\max}$ relative to the Kolmogorov and Batchelor length scales and the maximum time interval $(\Delta_{t})_{\max}$ relative to the Kolmogorov time scale. The complete set of simulation parameters is listed in table~\ref{tb:dns}. 
The Kolmogorov length scale is estimated as $\eta_{K} = (\nu^{3} / \langle \varepsilon_{u} \rangle_{V,t})^{1/4}$, while the Batchelor length scale is given by $\eta_{B} = \eta_{K} Pr^{-1/2}$. 
The Kolmogorov time scale is calculated as $\tau_{\eta} = \sqrt{\nu / \langle \varepsilon_{u} \rangle_{V,t}}$, where $\langle \varepsilon_{u} \rangle_{V,t}$ denotes the volume- and time-averaged turbulent kinetic energy dissipation rate, computed as $\langle \varepsilon_{u} \rangle_{V,t} = \nu \langle (\partial_{j} u'_{i})^{2} \rangle$, with $u'_{i}$ denoting the fluctuating velocity component. 
As summarised in table~\ref{tb:dns}, the maximum wall-normal grid spacing is of the same order as the Kolmogorov and Batchelor length scales, while the maximum time step remains substantially smaller than the Kolmogorov time scale in all cases. 
We also report the volume-averaged Nusselt number \(Nu\) from the present simulations alongside the reference values of \citet{Zhu2018PRL}, and find good agreement.
Details of the spectral element method and validation of the Nek5000 solver can be found in \citet{Fischer1997JCP} and \citet{Kooij2018CF}. As an additional verification, we also carried out simulations at $Ra = 10^{7}$ and $10^{8}$ using an in-house solver based on the lattice Boltzmann method \citep{Xu2017IJHMT}. The results from Nek5000 and the lattice Boltzmann solver show consistent flow fields and statistics.

\begin{table}
\centering
\begin{tabular}{ccccccc}
\toprule
$Ra$ & $N_x \times N_y$ & $(\Delta_g)_{\max}/\eta_K$ & $(\Delta_g)_{\max}/\eta_B$ & $(\Delta_t)_{\max}/\tau_\eta$ & $Nu$ (present) & $Nu$ \citep{Zhu2018PRL} \\
\midrule
$10^{7}$  & $1152 \times 288$  & 0.19 & 0.16 & 0.00021 & 14.5  & – \\
$10^{8}$  & $1152 \times 288$  & 0.97 & 0.81 & 0.00176 & 27.7  & 26.1 \\
$10^{9}$  & $2304 \times 576$  & 1.23 & 1.03 & 0.00137 & 48.0  & 48.3 \\
$10^{10}$ & $4608 \times 1152$ & 1.32 & 1.11 & 0.00069 & 95.1  & 95.1 \\
$10^{11}$ & $9216 \times 2304$ & 1.23 & 1.03 & 0.00037 & 185.2 & 193.6 \\
\bottomrule
\end{tabular} 
\caption{\textit{A posteriori} verification of spatial and temporal resolutions for the present direct numerical simulations. The columns list, from left to right: Rayleigh number \(Ra\); grid resolution \(N_x \times N_y\); maximum wall-normal grid spacing relative to the Kolmogorov length scale, \((\Delta_g)_{\max}/\eta_K\), and to the Batchelor length scale, \((\Delta_g)_{\max}/\eta_B\); maximum time step relative to the Kolmogorov time scale, \((\Delta_t)_{\max}/\tau_\eta\); volume-averaged Nusselt number \(Nu\) from the present simulations; and reference \(Nu\) values from \citet{Zhu2018PRL}.}
\label{tb:dns}
\end{table}

\subsection{Kinematics of self-propelled inertial particles}

We consider particles that are sufficiently small that they do not disturb the carrier flow. Here, `small' means that the particle diameter is smaller than the Kolmogorov length scale and larger than the molecular mean free path, so that Brownian motion is negligible. The particles are assumed to be isotropic, and in the present framework we consider only their translational motion. In a fluid environment with strong shear and vorticity, such as turbulent RB convection, particles are naturally subjected to flow-induced torques and rotational dynamics (e.g. Jeffery-type rotations). By restricting the model to translational kinematics, we effectively assume an idealised agent that either possesses an isotropic propulsion mechanism, capable of redirecting thrust without rotating its body, or has a negligible rotational relaxation time, thereby achieving instantaneous attitude control. Although this assumption allows us to isolate the higher-level trajectory-planning problem from the lower-level attitude-control problem, it also removes certain physical complexities. In particular, neglecting rotational dynamics means that the agent does not experience the continuous flow-induced torques that would otherwise perturb its orientation. Consequently, a real biological or artificial swimmer would likely incur an additional energetic cost to generate corrective torques against the local fluid vorticity in order to maintain its desired heading.

For the sub-Kolmogorov particles considered here, this additional rotational cost can be estimated by comparing the power required for rotational correction with that required for translational propulsion. For a small spherical particle, the translational power scales as \(P_{\rm trans}\sim 6\pi\mu aU^2\), while the rotational power scales as \(P_{\rm rot}\sim 8\pi\mu a^3\Omega^2\), where \(a\) is the particle radius, \(U\) is a characteristic relative translational velocity and \(\Omega\) is a characteristic relative angular velocity. Estimating the angular velocity from the local Kolmogorov-scale velocity gradient, \(\Omega\sim U/\eta_K\), gives \(P_{\rm rot}/P_{\rm trans}\sim (4/3)(a/\eta_K)^2\). In the present study, \(a=10~\mu{\rm m}\), while \(\eta_K\) is of the order of \(1.6\times10^{-3}\) to \(1.8\times10^{-3}\) m across the considered Rayleigh-number range. The resulting ratio is therefore \(P_{\rm rot}/P_{\rm trans}\sim O(10^{-5})\). This scaling estimate suggests that the rotational energy correction is small compared with the translational propulsion cost for the present sub-Kolmogorov regime, although rotational dynamics may no longer be negligible for larger, anisotropic or finite-size swimmers.

The total force acting on the particle consists of the net gravitational force $\boldsymbol{F}_{\rm gravity}$, the drag force $\boldsymbol{F}_{\rm drag}$ and the propulsive force $\boldsymbol{F}_{\rm propel}$.
In the present heavy-particle and sub-Kolmogorov regime \((\rho_p \gg \rho_f,\; St \sim 10^{-3})\), added-mass, undisturbed-flow pressure-gradient, shear-induced lift, and Basset history forces are expected to be small compared with the drag force and are therefore neglected.
The net gravitational force $\bm{F}_{\mathrm{gravity}}$, acting along the direction of gravitational acceleration after accounting for hydrostatic buoyancy, is given by

\begin{equation}
\bm{F}_{\mathrm{gravity}} = \rho_p V_p \bm{g} - \rho_f V_p \bm{g},
\end{equation}
where $\rho_p$ and $V_p$ denote the particle density and volume, respectively. The drag force $\bm{F}_{\mathrm{drag}}$ arises from the relative motion between the particle and the surrounding fluid and is given by

\begin{equation}
\bm{F}_{\mathrm{drag}} = \frac{m_p}{\tau_p} \left( \bm{u}_f - \bm{u}_p \right) f(Re_p),
\end{equation}
where $m_p$ and $\bm{u}_p$ denote the particle mass and velocity, respectively. The particle response time is $\tau_p = \rho_p d_p^2 / (18 \rho_f \nu)$, where $d_p$ is the particle diameter. The particle Reynolds number is $Re_p = d_p \|\bm{u}_f - \bm{u}_p \| / \nu$, which determines the drag correction factor $f(Re_p)$. 
For $Re_p \ll 1$, Stokes drag applies and $f \approx 1$; for finite $Re_p$, we use the Schiller--Naumann correction $f(Re_p) = 1 + 0.15\,Re_p^{0.687}$. The propulsive force $\bm{F}_{\mathrm{propel}} = m_p \bm{a}_{\mathrm{propel}}$ represents the particle's self-generated actuation and yields a propulsive acceleration bounded above by a prescribed limit. In natural systems, this actuation may arise from biochemical or thermal processes, whereas in artificial systems it may be driven by thermal, electrical or mechanical energy sources.

The particle motion is governed by

\begin{align}
\frac{d \bm{x}_p(t)}{dt} &= \bm{u}_p(t), \label{eq:xdim}\\[3pt]
\frac{d \bm{u}_p(t)}{dt} &= \frac{\rho_p - \rho_f}{\rho_p} \bm{g}
+ \frac{\bm{u}_f(\bm{x}_p, t) - \bm{u}_p(t)}{\tau_p} f(Re_p)
+ \bm{a}_{\mathrm{propel}}(t), \label{eq:udim}
\end{align}
where $\bm{x}_p$ denotes the particle position. 
After non-dimensionalisation using the scales introduced above, the governing equations become

\begin{align}
\frac{d \bm{x}_p^*}{dt^*} &= \bm{u}_p^*, \label{eq:xnd}\\[3pt]
\frac{d \bm{u}_p^*}{dt^*} &=
- \Lambda\,\hat{\bm{y}}
+ \frac{\bm{u}_f^* - \bm{u}_p^*}{St}\,f(Re_p)
+ \bm{a}_{\mathrm{propel}}^*, \label{eq:und}
\end{align}
where \(\Lambda=(\rho_p-\rho_f)/(\rho_p\beta\Delta_T)\) is the buoyancy ratio, which quantifies the particle buoyancy relative to the thermal buoyancy scale of the carrier flow.
The particle Stokes number is defined as $St = \tau_p / t_f$, where the free-fall time scale is $t_f = \sqrt{H / (g \beta \Delta_T)}$. 
The propulsion strength is characterised by the dimensionless activity number

\begin{equation}
\mathcal{A} = \frac{\|\bm{a}_{\mathrm{propel}}\|}{g},
\end{equation}
which measures the ratio of the particle's propulsive acceleration to gravity. The dimensionless propulsion term can therefore be written as

\begin{equation}
\bm{a}_{\mathrm{propel}}^*
= \mathcal{A} \frac{\rho_p}{\rho_p - \rho_f} \Lambda\,\hat{\bm{a}}_{\mathrm{propel}},
\end{equation}
where $\hat{\bm{a}}_{\mathrm{propel}} = \bm{a}_{\mathrm{propel}} / \|\bm{a}_{\mathrm{propel}}\|$ is the unit vector in the actuation direction. 
In the heavy-particle limit ($\rho_p \gg \rho_f$), $\rho_p / (\rho_p - \rho_f) \approx 1$, and the expression simplifies to

\begin{equation}
\bm{a}_{\mathrm{propel}}^* = \mathcal{A} \Lambda\,\hat{\bm{a}}_{\mathrm{propel}}.
\end{equation}
Accordingly, the dimensionless momentum equation reduces to

\begin{equation}
\frac{d \bm{u}_p^*}{dt^*}
= - \Lambda\,\hat{\bm{y}}
+ \frac{\bm{u}_f^* - \bm{u}_p^*}{St}\,f(Re_p)
+ \mathcal{A} \Lambda\,\hat{\bm{a}}_{\mathrm{propel}}.
\label{eq:und_final}
\end{equation}

\subsection{Optimal control of self-propelling particles} \label{sec:optimal}

We control the particle through its propulsive acceleration. The control input is given by

\begin{equation}
\bm{a}_{\mathrm{propel}} = \frac{\bm{F}_{\mathrm{propel}}}{m_p},
\end{equation}
with its magnitude bounded by \(\|\bm{a}_{\mathrm{propel}}\|/g = \mathcal{A} \leq \mathcal{A}_{\max}\), where \(\mathcal{A}_{\max}\) is the prescribed actuation limit. At each decision step, the self-propelled particle serves as the agent in an RL framework. It receives an observation of the local flow state, selects a propulsive acceleration and then receives a reward together with the next observation. Through repeated interactions, the agent seeks a policy that maximises the expected cumulative return \citep{Mehta2019PR, Brunton2020ARFM}.

Reinforcement-learning methods are commonly classified as either policy-based or value-based. Policy-based methods, such as policy-gradient methods, optimise the parameters \(\bm{\theta}\) of a stochastic policy to maximise a performance objective \(J(\pi_{\bm{\theta}})\). They naturally accommodate continuous actions, but can require many samples and may converge slowly, making them less suitable for complex flow problems. By contrast, value-based methods, such as \(Q\)-learning, learn an action--value function \(Q_{\bm{\theta}}(s,a)\) and derive a deterministic policy by selecting the action that maximises it, i.e. \(a(s)=\arg\max_{a} Q_{\bm{\theta}}(s,a)\). They are efficient for discrete actions but are generally not well suited to continuous control.

For navigation problems with continuous state and action spaces, the soft actor--critic (SAC) method is well suited because it combines policy-based and value-based components. The actor network selects the propulsive acceleration, while the critic network evaluates the corresponding value function to guide learning. The SAC algorithm maximises not only the expected cumulative reward but also the policy entropy, thereby encouraging exploration. 
The optimal policy $\pi^*$ is obtained by solving

\begin{equation}
\pi^* = \arg\max_{\pi} \mathbb{E}_{\tau \sim \pi} \left[ \sum_{t=0}^{\infty} \gamma^t \left\{ r_t(s_t, a_t, s_{t+1}) + \alpha \mathcal{H}[\pi(\cdot|s_t)] \right\} \right],
\end{equation}
where $\gamma$ is the discount factor, and the coefficient $\alpha$ controls the trade-off between the entropy term and the reward. The
reward $r_t$ depends on the state $s_t$, the chosen action $a_t$ and the resulting transition to $s_{t+1}$.
The entropy $\mathcal{H}$ of the policy at state $s_t$ measures the randomness of the selected actions and
is computed from the action distribution as $\mathcal{H}[\pi(\cdot|s_t)] = \mathbb{E}_{a_t \sim \pi(\cdot|s_t)} [-\log \pi(a_t|s_t)]$.
The SAC algorithm therefore seeks a policy that maximises a weighted sum of expected reward and entropy, enabling efficient exploration while maintaining training stability \citep{Haarnoja2018ICML}. 
Detailed descriptions of the SAC formulation are provided in Appendix~\ref{sec:appSAC}.

Key components of the RL framework include the environmental cues available to the agent (i.e. the current state \(s_t\)), the agent's actions (i.e. \(a_t\)) and the feedback it receives for its behaviour (i.e. the reward \(r_t\)). The observation at time \(t^*\) is given by

\begin{equation}
s_t = \big\{ \bm{x}_p^*, \bm{u}_p^*, \bm{a}_p^*, \bm{u}_f^*, \nabla \bm{u}_f^*, T_f^*, (\nabla T_f)_{x}^* \big\},
\end{equation}
where $\nabla \bm{u}_f^*$ denotes the local velocity gradient and $(\nabla T_f)_{x}^*$ the local horizontal temperature gradient. 
We emphasise that the agent relies on locally measurable environmental cues evaluated at its instantaneous position, rather than on any globally available description of the flow field. 
Although sensing the absolute location $\bm{x}_p$ is typically unfeasible for microscopic swimmers, it is readily available for macroscopic autonomous vehicles through standard positioning systems. 
Recent studies have cautioned that including absolute coordinates may allow agents to artificially `memorise' static flow signatures in steady or periodic flows \citep{Gunnarson2021NC,Jiao2025Gradients}. 
However, at the moderate to high Rayleigh numbers considered here, thermal convection is spatiotemporally chaotic, and the locations of plumes and coherent structures continuously drift and evolve. This inherent unsteadiness reduces the likelihood that the agent simply overfits to absolute coordinates. We therefore retain the spatial information \(\bm{x}_p\) because the task involves a fixed-displacement target and the agent requires global context to assess its progress. Furthermore, unlike in our previous study at lower \(Ra\) \citep{Xu2023PRF}, the inclusion of local gradients \((\nabla \bm{u}_f,(\nabla T_f)_x)\) becomes important at high Rayleigh numbers owing to the emergence of finer characteristic scales. The information carried by \((\nabla T_f)_x\) cannot be inferred solely from the instantaneous velocity gradient \(\nabla \bm{u}_f\). Whereas \(\nabla \bm{u}_f\) encodes local shear and strain, \((\nabla T_f)_x\) is related to vorticity production in two-dimensional thermal convection. It therefore provides the agent with additional information about local flow acceleration, a physical mechanism specific to active-scalar turbulence.

A systematic ablation study examining the role and non-redundancy of these positional and gradient-based observations is provided in Appendix~\ref{sec:appObservation}. Briefly, the ablation results show that the relative importance of different sensory inputs depends on the flow regime. At moderate \(Ra\), for example \(Ra=10^8\), the flow remains relatively organised by LSC, and local gradient cues are correlated with the large-scale transport structure, leading to a clear improvement in navigation performance. At higher \(Ra\), for example \(Ra=10^{10}\), the flow becomes more intermittent and less spatially coherent. In this regime, explicit positional information \(\boldsymbol{x}_p\) becomes important for maintaining global progress towards the fixed-displacement target. The ablation study also shows that, at high \(Ra\), providing the agent with local velocity-gradient information \(\nabla \boldsymbol{u}_f\) yields better navigation performance than providing the local temperature field \(T\) alone. A possible physical explanation is that thermal plumes become increasingly fragmented and intermittent at high \(Ra\), so the scalar temperature field may become less directly predictive of subsequent transport. By contrast, the velocity-gradient tensor encodes local kinematic topology, including shear layers and vortex boundaries through quantities such as the \(Q\)-criterion. It therefore provides the agent with local cues that help it avoid trapping in vortical regions and access strain-dominated transport pathways.

The action corresponds to the particle's propulsive acceleration. We parametrise it as a bounded continuous action in polar form,

\begin{equation}
\bm{a}_{\mathrm{propel}} = \mathcal{A}_{\max} g [c_a \cos(\theta),\, c_a \sin(\theta)],
\end{equation}
with $c_a = (a_0 + 1)/2 \in [0,1]$ and $\theta = \pi a_1 / 2 \in [-\pi/2, \pi/2]$. 
This half-plane restriction reflects the physical constraint that the particle cannot generate reverse thrust. 
To assess the robustness of this choice, we conducted sensitivity checks for representative cases, $Ra=10^8$ and $10^{10}$, by evaluating agents with a full action space, $\theta\in[-\pi,\pi]$. 
These unconstrained agents learned to avoid backward thrust in order to maximise the reward, yielding success-rate curves nearly identical to those of the constrained cases.
The variables $a_0$ and $a_1$ are the normalised outputs of the policy and lie in $[-1,1]$. 
The realised dimensionless actuation magnitude therefore satisfies $\|\bm{a}_{\mathrm{propel}}\|/g = \mathcal{A} = \mathcal{A}_{\max} c_a  \le \mathcal{A}_{\max}$.

The reward function balances progress towards the target against the cost of propulsion,

\begin{equation}
r_t = \frac{1}{R+Q} \big[ R\,V_{\mathrm{eff}}^{*}(t) - Q\,\|\bm{a}_{\mathrm{propel}}^{*}(t)\| \big],
\end{equation}
where \(V_{\mathrm{eff}}^{*}(t) = \bm{u}_p^{*}(t) \cdot \bm{n}\) is the dimensionless particle velocity along the target direction \(\bm{n}\). The first term promotes rapid motion towards the goal, whereas the second penalises excessive propulsive effort. The coefficients \(R\) and \(Q\) determine the trade-off between completion time and energy expenditure. The normalisation factor \(1/(R+Q)\) renders the reward a normalised weighted sum of the two physical objectives. This helps ensure that the magnitude of the single-step reward remains bounded by the characteristic physical scales, namely the dimensionless velocity \(V_{\mathrm{eff}}^{*}\) and the actuation limit \(\mathcal{A}_{\max}\), thereby reducing the risk of arbitrary inflation of the reward signal and unstable gradients during training for different choices of \(R\) and \(Q\). To guide the reader through the subsequent analyses, we note that in the primary evaluations presented in \S~\ref{sec:ResultsDiscussion} we fix the reward-weighting ratio at a large value of \(R/Q=5000\). As shown in \S~\ref{sec:trade-offs}, a sufficiently large \(R/Q\) ratio is useful in highly chaotic RB convection. It helps the agent prioritise reachability and time efficiency, which are required to cross transport barriers between convective rolls, while the energy-penalty term simultaneously discourages redundant or purely random actuation. A detailed parametric investigation of this trade-off, including the limiting behaviours of time-optimal and energy-optimal policies, is deferred to \S~\ref{sec:trade-offs}.

\subsection{Simulation settings}

We study the motion of self-propelled particles in a two-dimensional convection cell of size $L \times H$, with aspect ratio $\Gamma = L / H = 4$. 
The top and bottom boundaries are maintained at constant cold and hot temperatures, respectively, and satisfy no-slip velocity boundary conditions. 
The left and right boundaries are periodic. Simulations are performed over the Rayleigh-number range $10^{7} \le Ra \le 10^{11}$ at fixed Prandtl number $Pr = 0.71$. Because modelling self-propelled particles introduces more parameters than passive tracers and each simulation is computationally expensive, an exhaustive exploration of parameter space is not practical. Unless stated otherwise, we therefore focus on the effect of the maximum propulsive acceleration, $\mathcal{A}_{\max} = \max(\mathcal{A}) = \max(\|\bm{a}_{\mathrm{propel}}\|/g) \in [0,5]$, which characterises the particle's actuation limit. The particle diameter and density are fixed at $d_p = 20~\mu\mathrm{m}$ and $\rho_p = 3000~\mathrm{kg\,m^{-3}}$, respectively; over the Rayleigh-number range considered, the corresponding Stokes number $St$ varies from $5.52 \times 10^{-3}$ at $Ra = 10^{7}$ to $1.19 \times 10^{-3}$ at $Ra = 10^{11}$.

We consider a fixed-displacement task. Throughout training, the self-propelled particle serves as the agent and is trained for a fixed budget of \(1\times 10^6\) training steps without early stopping. Its task is to achieve a displacement \(\ell=[\bm{x}(t)-\bm{x}(0)]\cdot\bm{n}\) along the direction \(\bm{n}\). For simplicity, we set \(\bm{n}=(1,0)\) and \(\ell=4\). Because length is non-dimensionalised by the cell height \(H\), this target distance corresponds to traversing one full domain width of \(L/H=4\). An episode is counted as successful as soon as the particle achieves a net horizontal displacement of 4, regardless of its final vertical position. This choice is motivated by the lateral traversal of autonomous vehicles in atmospheric and oceanic environments. Our computational domain is intentionally anisotropic to reflect this setting. Specifically, it is periodic in the horizontal direction but bounded by solid walls in the vertical direction. Consequently, horizontal and vertical motions are inequivalent. The horizontal task allows the particle to traverse the domain continuously, thereby focusing on the challenge of crossing transport barriers (e.g. roll edges) and navigating intermittent cross-flows. An episode terminates unsuccessfully if the particle settles on the bottom wall or if the maximum allowed time, defined as \(t^*_{\max}=30\), is reached. As shown in \S~\ref{sec:performance}, successful traversals complete within \(15 \leq t^* \leq 20\). Thus, \(t^*_{\max}=30\) provides a sufficient margin, corresponding to \(1.5\)--\(2.0\) times the typical completion time. Sensitivity tests conducted for representative Rayleigh numbers, namely \(Ra=10^8\) and \(10^{10}\), in which the time limit was extended to \(t^*_{\max}=60\), yielded unchanged success rates. This suggests that failures due to the time limit primarily reflect particles becoming trapped in recirculating vortex cores because of insufficient propulsive thrust, rather than slower trajectories that would otherwise succeed being misclassified as failures.

\subsection{Comparative constant-heading baseline} \label{sec:appNaive}

To provide a benchmark for propulsion-energy consumption, we construct a constant-heading baseline. This baseline acts as an inverse-dynamics trajectory-tracking controller. Specifically, it uses the completion time \(t_{\text{RL}}\) achieved by the RL policy to prescribe uniform horizontal progress across the domain. Although this gives the baseline privileged kinematic information, its role is purely comparative. By requiring the baseline to complete the traversal in the same total time as the RL agent, we obtain a controlled comparison of propulsion-energy consumption under an identical kinematic time constraint. The baseline therefore uses the same total duration and prescribes equal horizontal progress during each control interval of length \(\Delta t_c\). For the target direction \(\bm{n}=(1,0)\), the dimensional desired displacement over one control interval is

\begin{equation}
\Delta \bm{x}_{\text{des}} = \frac{\ell}{t_{\text{RL}}}\Delta t_{c}\bm{n} = \left(\frac{\ell \Delta t_{c}}{t_{\text{RL}}}, 0\right),
\label{eq:C1}
\end{equation}
where the desired vertical displacement is zero.

Assuming constant particle acceleration over each control interval $\Delta t_c$, the displacement satisfies

\begin{equation}
\Delta \bm{x} = \bm{u}_p \Delta t_c + \tfrac{1}{2}\bm{a}_p \Delta t_c^2,
\end{equation}
where $\bm{u}_p$ is the particle velocity at the start of the interval. The corresponding total force is therefore

\begin{equation}
\bm{F}_{\text{total}} = m_p \bm{a}_p = \frac{2m_p}{(\Delta t_c)^2}\left(\Delta \bm{x}_{\text{des}} - \bm{u}_p \Delta t_c\right).
\label{eq:C2}
\end{equation}
Using the gravitational and drag forces defined above, the required propulsive force is

\begin{equation}
\bm{F}_{\text{propel}}(t) = \bm{F}_{\text{total}}(t) - \bm{F}_{\text{gravity}}(t) - \bm{F}_{\text{drag}}(t).
\label{eq:C3}
\end{equation}

To ensure a fair comparison, the baseline agent operates with the same control interval \(\Delta t_c\) as the learned RL policy. The particle applies \(\bm{F}_{\text{propel}}\) over one control interval \(\Delta t_c\), which contains \(N_{\text{sub}}\) particle-integration substeps of size \(\Delta t_p\), and we set \(\Delta t_c = 30\,\Delta t_p\). For the RL agent, this ratio represents a practical compromise between learning efficiency and physical control responsiveness. For numerical stability, the dimensionless particle-integration time step must remain small (e.g. \(\Delta t_p \sim 10^{-3} t^*\)). If the agent were to make decisions at every integration step (\(\Delta t_c=\Delta t_p\)), a single episode would contain \(\mathcal{O}(10^4)\) training steps. Such long horizons make the RL credit-assignment process difficult and can impede convergence. Conversely, choosing \(\Delta t_c\) too large would prevent the agent from responding to rapidly evolving transient structures in the convective flow. The choice \(\Delta t_c=30\,\Delta t_p\) gives a complete trajectory involving \(\mathcal{O}(10^2)\) decision steps, keeping training computationally tractable while maintaining sufficient control frequency to navigate intermittent turbulence. The baseline agent adopts the same interval in order to match the decision frequency of the RL agent.

Because the turbulent background flow perturbs the actual trajectory, after each interval we update the remaining time and the remaining horizontal displacement, \(\ell_{\text{rem}}\), projected along \(\bm{n}\). The target increment for the next interval is then given by

\begin{equation}
\Delta \bm{x}_{\text{des}} = \frac{\ell_{\text{rem}}}{t_{\text{rem}}}\Delta t_{c}\bm{n},
\label{eq:C6}
\end{equation}
and (\ref{eq:C2})--(\ref{eq:C6}) are reapplied until the particle reaches the target or the maximum allowed time is reached.

\section{Results and discussion}\label{sec:ResultsDiscussion}

\subsection{Flow organisation and macroscopic navigation performance} \label{sec:performance}

\begin{figure}
\centering
\includegraphics[width=0.5\textwidth]{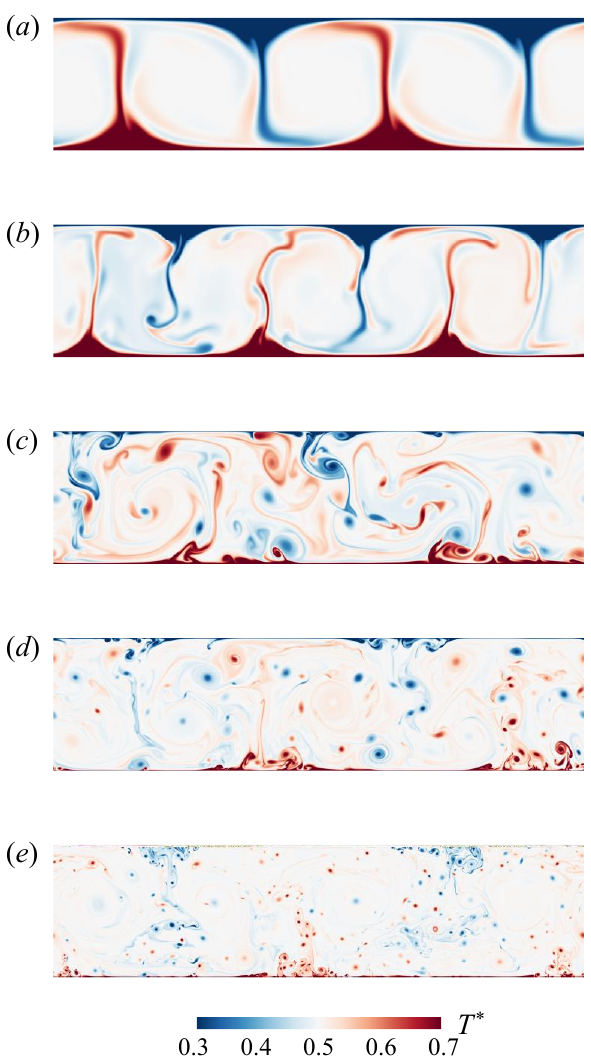}
\caption{Instantaneous dimensionless temperature field \(T^*\) at \(Pr=0.71\) in a cell of aspect ratio \(\Gamma=4\) for (\textit{a}) \(Ra=10^7\), (\textit{b}) \(Ra=10^8\), (\textit{c}) \(Ra=10^9\), (\textit{d}) \(Ra=10^{10}\) and (\textit{e}) \(Ra=10^{11}\). Here, \(T^*=(T-T_0)/\Delta_T\), where \(\Delta_T\) denotes the temperature difference between the heated bottom and cooled top walls.}
\label{fig:instant_temperature}
\end{figure}

Figure~\ref{fig:instant_temperature} shows instantaneous temperature contours at several values of \(Ra\). At \(Ra=10^7\) (see figure~\ref{fig:instant_temperature}\textit{a}), a four-roll LSC spans the domain. Thin thermal boundary layers develop along the horizontal walls, while the roll interiors remain relatively quiescent. At \(Ra=10^8\) (see figure~\ref{fig:instant_temperature}\textit{b}), the roll structure persists, but the number of rolls increases from four to six; see the discussion in \citet{Wang2020PRL} and \citet{Xu2023PRF}. Small-scale instabilities develop, producing wavy shear layers and secondary eddies near the corners and roll edges, and plume detachment becomes more frequent. At \(Ra=10^9\) (see figure~\ref{fig:instant_temperature}\textit{c}), the LSC weakens, and the interior contains smaller, less coherent vortices. Plume ejection and impingement become more frequent and exhibit greater spatial variability. At \(Ra=10^{10}\) (see figure~\ref{fig:instant_temperature}\textit{d}), additional small-scale vortices emerge, and the bulk temperature field becomes well mixed, although regions of hot-plume ejection and cold-plume impingement remain identifiable. At \(Ra=10^{11}\) (see figure~\ref{fig:instant_temperature}\textit{e}), the flow is dominated by small scales and is mixed throughout the bulk. A sustained multi-roll LSC is no longer evident, but plume activity near the horizontal boundaries remains pronounced. Overall, the flow patterns in the \(\Gamma=4\) cell are consistent with those reported for the \(\Gamma=2\) cell by \citet{Zhu2018PRL} and for the \(\Gamma=1\) cell by \citet{Zhang2017JFM}, \citet{he2024turbulent} and \citet{Gao2024JFM}. A detailed comparison of the flow statistics in terms of the Nusselt number is provided in table~\ref{tb:dns}. For reviews of LSC and plume dynamics, see \citet{Lohse2010ARFM}, \citet{ Chilla2012EPJE}, \citet{Xia2023NSR}, \citet{Lohse2023PT,Lohse2024RMP} and \citet{Xia2025ActaMS}.

\begin{figure}
\centering
\includegraphics[width=\textwidth]{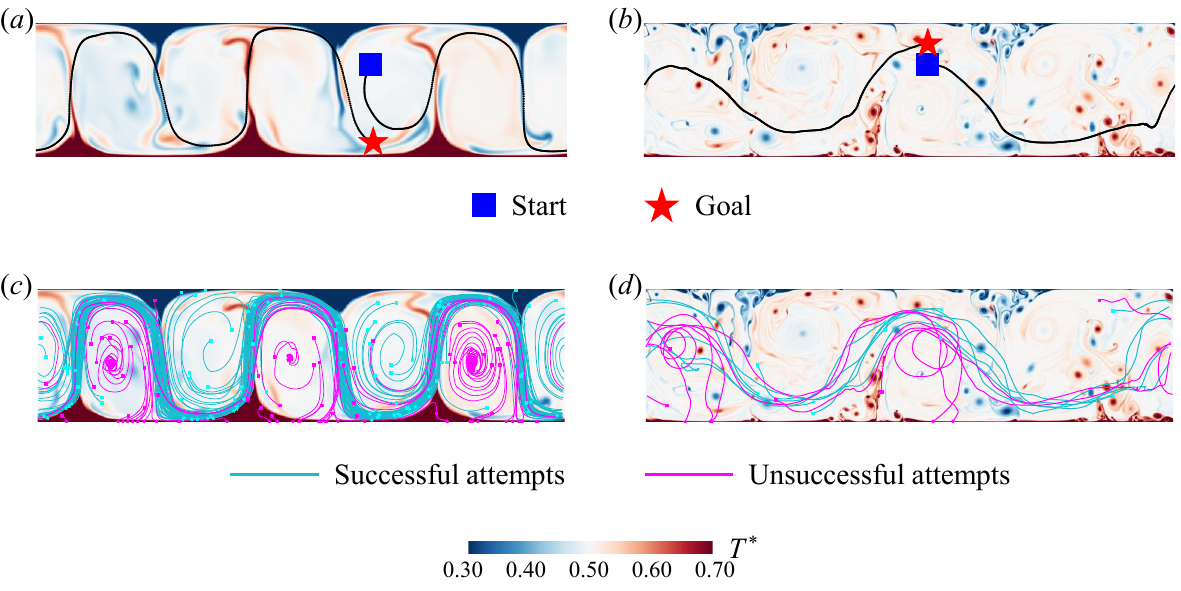}
\caption{Representative navigation trajectories for the fixed-displacement task. 
(\textit{a},\textit{b}) Representative single trajectories and  (\textit{c},\textit{d}) trajectory ensembles, for (\textit{a}) \(Ra=10^8\) with \(\mathcal{A}_{\max}=1.5\), (\textit{b}) \(Ra=10^{10}\) with \(\mathcal{A}_{\max}=5.0\), (\textit{c}) \(Ra=10^8\) with \(\mathcal{A}_{\max}=1.0\) and (\textit{d}) \(Ra=10^{10}\) with \(\mathcal{A}_{\max}=3.0\). The background contours indicate the instantaneous dimensionless temperature \(T^*\). The prescribed horizontal displacement is \(\ell=4\), equal to one full domain width (\(\Gamma=4\)), while the final vertical position is unconstrained. Because the lateral boundaries are periodic, the particle traverses the domain once. As a result, the start (blue square) and goal (red star) markers have the same horizontal coordinate in the plotted window; any apparent offset is due to the unconstrained vertical drift.}
\label{fig:navigation_trajectories}
\end{figure}

Figure~\ref{fig:navigation_trajectories} shows representative trajectories for the fixed-displacement task. The particle starts from the blue square and is required to reach a target displaced by \(\ell\,\bm{n}\). Here we set \(\bm{n}=(1,0)\) and \(\ell=4\), so the target, marked by a red star, lies one domain width to the right in a cell of aspect ratio \(\Gamma=4\). Because the lateral boundaries are periodic, a trajectory that exits one lateral boundary re-enters from the opposite side and continues rightward. At \(Ra=10^8\) and \(\mathcal{A}_{\max}=1.5\) (see figure~\ref{fig:navigation_trajectories}\textit{a}), the trajectory largely follows the roll edges \citep{Emran2010PRE}. At \(Ra=10^{10}\) and \(\mathcal{A}_{\max}=5.0\) (see figure~\ref{fig:navigation_trajectories}\textit{b}), the trajectory becomes more irregular, reflecting enhanced small-scale activity. The particle advances in short, plume-assisted segments and occasionally crosses between rolls. Figures~\ref{fig:navigation_trajectories}(\textit{c}) and \ref{fig:navigation_trajectories}(\textit{d}) further show ensembles of particle trajectories obtained with the bounded-acceleration policy. Successful attempts are shown in cyan, and unsuccessful attempts in magenta. At \(Ra=10^8\), unsuccessful trajectories are trapped within recirculating cores before reaching the time limit \(t_{\max}^{*}\) or hitting the bottom wall, whereas successful trajectories follow narrow pathways near roll edges. At \(Ra=10^{10}\), confinement weakens and trajectories become more space-filling; several successful examples exploit transient plume-assisted pathways to reach the target. These results suggest that increasing turbulent activity weakens coherent confinement and broadens the available pathways, thereby improving reachability under a fixed actuation limit.

\begin{figure}
\centering
\includegraphics[width=0.9\textwidth]{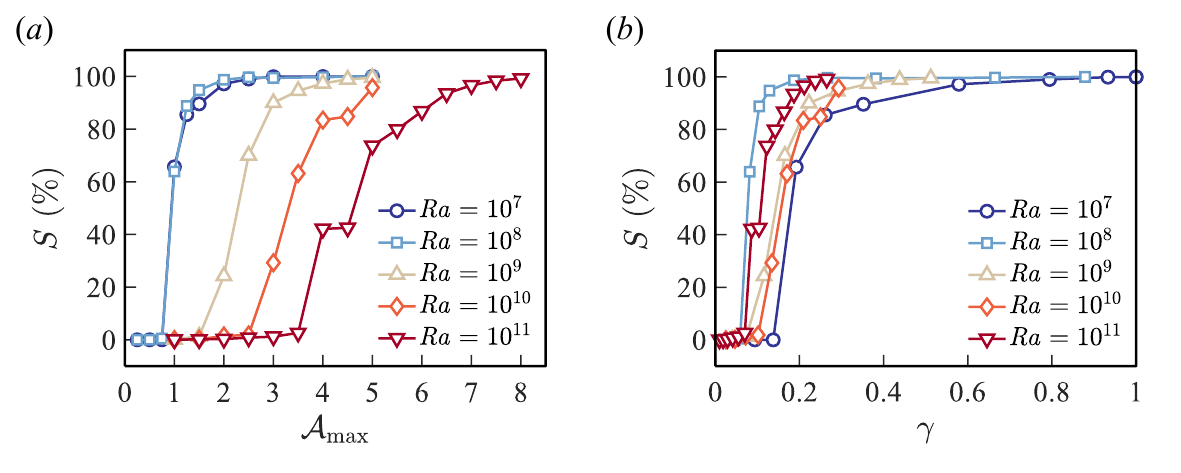}
  \caption{Success rate \(S=N_1/N_0\) as a function of (\textit{a}) the maximum dimensionless propulsive acceleration, \(\mathcal{A}_{\max}=\max(\|\boldsymbol{a}_{\mathrm{propel}}\|/g)\), and (\textit{b}) the navigable-area fraction, \(\gamma\), for Rayleigh numbers \(10^7 \le Ra \le 10^{11}\). Here, \(\gamma\) denotes the spatiotemporal mean fraction of the flow domain in which the particle's active terminal velocity \(U_{t,\max}^*\) exceeds the local fluid speed \(\|\boldsymbol{u}_f^*\|\).}
\label{fig:success_rate}
\end{figure}

Having described the qualitative differences in particle trajectories across flow regimes, we now quantify the macroscopic navigation performance in terms of reachability, completion time and energetic cost. In the following, the dimensionless maximum propulsive acceleration, \(\mathcal{A}_{\max}\), is used to characterise the actuation limit of the self-propelled particle.
Figure~\ref{fig:success_rate}(\textit{a}) shows the success probability, estimated empirically as the success rate \(S=N_1/N_0\), as a function of \(\mathcal{A}_{\max}\) for various \(Ra\). Each curve is based on \(N_0=10^4\) uniformly distributed particle releases under identical flow and boundary conditions, and \(N_1\) denotes the number of trials that complete the fixed-displacement task. The success rate increases monotonically with \(\mathcal{A}_{\max}\) for all \(Ra\), but the shape of the \(S\) curve varies with \(Ra\). At moderate Rayleigh numbers (\(Ra=10^7\) and \(10^8\)), the curves exhibit a sharp transition. Specifically, \(S\) remains zero for \(\mathcal{A}_{\max}\lesssim 1.0\), then rises steeply to about \(0.8\) over a narrow interval and approaches unity for \(\mathcal{A}_{\max}\gtrsim 1.5\). This behaviour suggests a finite acceleration threshold required for the particle to cross the boundaries between neighbouring large-scale rolls \citep{Haller2015ARFM, Schneide2019PRE}. As \(Ra\) increases to \(10^9\) and \(10^{10}\), the transition shifts to larger \(\mathcal{A}_{\max}\) and becomes more gradual. At \(Ra=10^{11}\), the success rate remains low (about \(30\%\)) for \(\mathcal{A}_{\max}\lesssim 3\) and reaches about \(80\%\) only when \(\mathcal{A}_{\max}\gtrsim 4\). The rightward shift and progressive smoothing of the transition with increasing \(Ra\) are consistent with the weakening of coherent roll barriers and the strengthening of intermittent motions, which require stronger and more sustained propulsive acceleration to keep trajectories aligned with the target direction.

To examine the relation between the rightward shift and the increasing flow intensity at higher \(Ra\), we compare the agent's kinematic capability with the Eulerian flow statistics. We define the particle's dimensionless active terminal velocity in a quiescent fluid, determined by the balance between horizontal propulsion and Stokes drag, as \(U_{t,\max}^* = St\mathcal{A}_{\max}\Lambda\). We then calculate the navigable-area fraction \(\gamma\), defined as the spatiotemporal average fraction of the domain in which the particle's terminal speed exceeds the local fluid speed, i.e. \(U_{t,\max}^* > \|\bm{u}_{f}^*\|\). As shown in figure~\ref{fig:success_rate}(\textit{b}), plotting the success rate against the navigable-area fraction \(\gamma\) largely collapses the curves for all Rayleigh numbers. The success rate exhibits a sharp increase over a narrow interval, \(0.05 \le \gamma \le 0.20\). This result suggests that the higher actuation required at large \(Ra\) is predominantly associated with the shrinking fraction of kinematically navigable regions. Thus, \(\gamma\) provides a useful reduced parameter for organising the baseline reachability data within the present convective-flow setting. However, \(\gamma\) alone does not fully determine the success probability, because it measures local kinematic accessibility but does not encode the finite-time connectivity of the navigable regions. 
Different flows, or different time windows, with similar \(\gamma\) may therefore yield different success probabilities if their LCS connectivity differs. The slight residual spread among the collapsed curves in figure~\ref{fig:success_rate}(\textit{b}) is likely related to this effect. In particular, as \(Ra\) increases, coherent roll barriers become increasingly fragmented and intermittent plume-assisted pathways emerge, modifying the transport connectivity even at comparable values of \(\gamma\).

\begin{figure}
\centering
\includegraphics[width=0.9\textwidth]{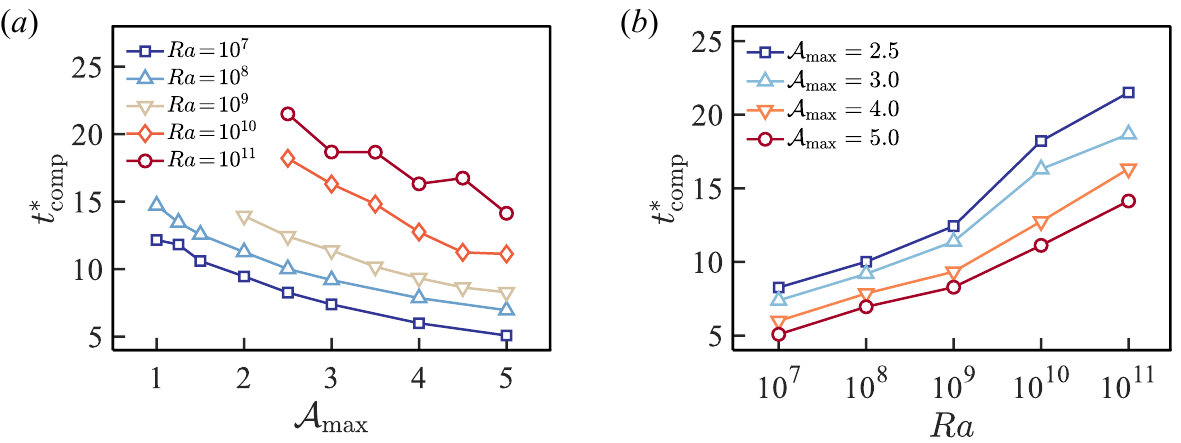}
  \caption{Dimensionless completion time \(t_{\mathrm{comp}}^*\) for the fixed-displacement task as a function of (\textit{a}) the maximum dimensionless propulsive acceleration \(\mathcal{A}_{\max}\) for different \(Ra\) and (\textit{b}) the Rayleigh number \(Ra\) for different \(\mathcal{A}_{\max}\).}
\label{fig:completion_time}
\end{figure}

We next examine the dimensionless completion time \(t^*_{\text{comp}}\) for the fixed-displacement task. Figure~\ref{fig:completion_time}(\textit{a}) shows that \(t^*_{\text{comp}}\) decreases monotonically with the maximum propulsive acceleration \(\mathcal{A}_{\max}\) for all Rayleigh numbers. The strongest reduction occurs for \(1 \lesssim \mathcal{A}_{\max} \lesssim 3\); beyond this range, the curves gradually flatten, indicating diminishing time savings as the acceleration is increased further. This behaviour is consistent with the presence of large-strain boundaries and plume channels that act as finite-time transport barriers. Once the applied acceleration is sufficient to cross these barriers, further increases in \(\mathcal{A}_{\max}\) yield only modest gains \citep{Haller2015ARFM, Schneide2019PRE}. Figure~\ref{fig:completion_time}(\textit{b}) provides a complementary perspective. For each actuation level, \(t^*_{\text{comp}}\) increases approximately monotonically with \(Ra\). The separation between the curves widens with \(Ra\), but their ordering with respect to \(\mathcal{A}_{\max}\) remains unchanged. Larger \(\mathcal{A}_{\max}\) consistently corresponds to shorter \(t^*_{\text{comp}}\) at a given \(Ra\).

\begin{figure}
\centering
\includegraphics[width=0.9\textwidth]{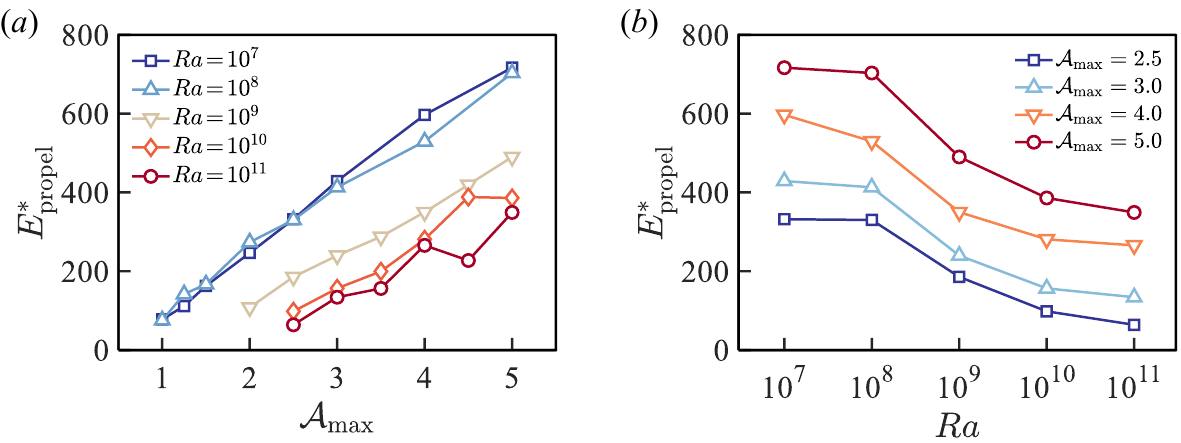}
  \caption{Dimensionless propulsion energy \(E_{\mathrm{propel}}^*\) for the fixed-displacement task as a function of (\textit{a}) the maximum dimensionless propulsive acceleration \(\mathcal{A}_{\max}\) for different \(Ra\) and (\textit{b}) the Rayleigh number \(Ra\) for different \(\mathcal{A}_{\max}\).}
\label{fig:propulsion_energy}
\end{figure}

We next examine the dimensionless propulsion energy \(E_{\mathrm{propel}}^*\) for the fixed-displacement task. 
To enable a meaningful comparison across parameter values, the dimensional energy \(E_{\mathrm{propel}} = \int \boldsymbol{F}_{\mathrm{propel}} \cdot \mathrm{d}\boldsymbol{s}\) is non-dimensionalised by the reference mechanical work \(m_p g \beta \Delta_T H\), giving

\begin{equation}
    E_{\mathrm{propel}}^* = \frac{E_{\mathrm{propel}}}{m_p g \beta \Delta_T H} = \int \boldsymbol{a}_{\mathrm{propel}}^* \cdot \mathrm{d}\boldsymbol{s}^*,
    \label{eq:energy_dimensionless}
\end{equation}
where \(\boldsymbol{s}^* = \boldsymbol{s}/H\) is the dimensionless path coordinate. 
The integral is evaluated only for successful trajectories and then averaged over them.
Figure~\ref{fig:propulsion_energy}(\textit{a}) plots \(E_{\mathrm{propel}}^*\) versus the maximum dimensionless propulsive acceleration \(\mathcal{A}_{\max}\) for different \(Ra\). 
For all \(Ra\), \(E_{\mathrm{propel}}^*\) increases with \(\mathcal{A}_{\max}\). 
At moderate \(Ra = 10^{7}\) and \(10^{8}\), the increase is nearly linear and the curves have similar slopes.
This linear scaling is consistent with traversal across successive coherent LSC cells. 
In this regime, each increment in actuation produces a comparable increment in mechanical work, as also reported for active-particle transport in laminar or weakly chaotic flows \citep{Piro2024PRR}. 
As \(Ra\) rises to \(10^{9}\) and above, the curves bend downward and shift to lower values. 
For a given \(\mathcal{A}_{\max}\), the required propulsion energy is smaller and less sensitive to \(\mathcal{A}_{\max}\), suggesting a larger contribution from background plume motions and small-scale intermittency. 
Similar reductions in propulsion cost with increasing environmental fluctuations have been reported for microswimmers in turbulence and active suspensions \citep{Yang2019PRF}. 
Figure~\ref{fig:propulsion_energy}(\textit{b}) shows \(E_{\mathrm{propel}}^*\) versus \(Ra\) at fixed \(\mathcal{A}_{\max}\). 
For each \(\mathcal{A}_{\max}\), \(E_{\mathrm{propel}}^*\) generally decreases with \(Ra\), with only mild non-monotonic variations. 
The narrowing of the spacing between curves at high \(Ra\) suggests that the dependence of \(E_{\mathrm{propel}}^*\) on \(\mathcal{A}_{\max}\) weakens; further increases in \(\mathcal{A}_{\max}\) then produce only small changes in \(E_{\mathrm{propel}}^*\). 
Taken together with the success-rate trends, these results suggest that more vigorous flows at higher \(Ra\) require larger \(\mathcal{A}_{\max}\) to achieve reachability, but that once reachability is attained, the energetic cost of traversal decreases. 
This trade-off between flow-assisted transport and actuation cost is consistent with observations for active agents in complex flows \citep{Ishikawa2025JFM}.

\subsection{Flow coherence and policy generalisability}

To further examine the \(Ra\) dependence of the success-rate curves, we analyse the organisation of the carrier flow. Specifically, we apply POD to extract the dominant coherent structures. The POD method has been widely used to study LSC dynamics in convection cells \citep{Podvin2015JFM, Castillo2019JFM, Soucasse2019JFM}. The spatiotemporal velocity field \(\bm{u}(\bm{x},t)\) is expressed as a superposition of empirical orthogonal eigenfunctions \(\bm{\phi}_i(\bm{x})\) with time-dependent amplitudes \(a_i(t)\):

\begin{equation}
\bm{u}(\bm{x},t) = \sum_{i=1}^{\infty} a_i(t)\, \bm{\phi}_i(\bm{x}).
\end{equation}
Here, \(\bm{u}(\bm{x},t)=[u(\bm{x},t),\,v(\bm{x},t)]^{T}\) is the two-dimensional velocity field, \(\bm{\phi}_i(\bm{x})=[\phi_{i}^{u}(\bm{x}),\,\phi_{i}^{v}(\bm{x})]^{T}\) are the spatial eigenfunctions (the POD modes) and \(a_i(t)\) are the temporal coefficients. Each orthonormal spatial mode \(\bm{\phi}_i(\bm{x})\) has an eigenvalue \(\lambda_i\) representing the fraction of fluctuation energy captured by mode \(i\). The eigenvalues are normalised such that \(\sum_i \lambda_i = 1\). We report the cumulative energy \(\sum_{i=1}^{m}\lambda_i\) as a function of the number of retained modes \(m\), which indicates how many modes are required to represent the dominant dynamics. Rapid saturation of this curve indicates strong flow coherence, whereas slow saturation implies that the energy is distributed over many modes.

\begin{figure}
\centering
\includegraphics[width=\textwidth]{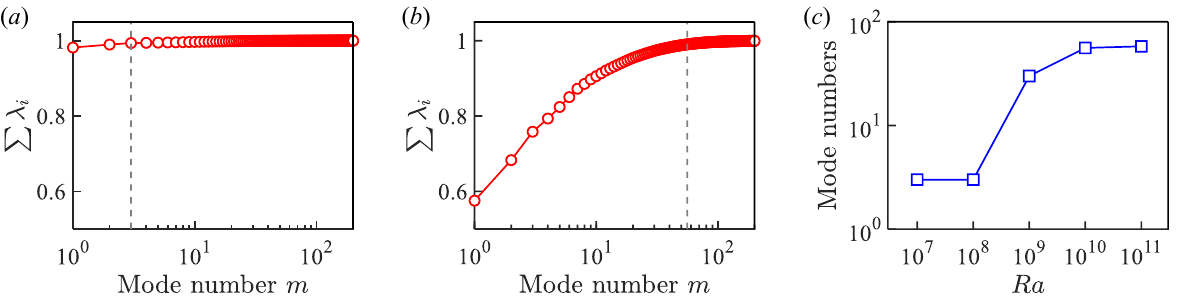}
  \caption{Normalised cumulative modal energy \(\sum_{i=1}^m \lambda_i\) as a function of mode number \(m\) for (\textit{a}) \(Ra=10^8\) and (\textit{b}) \(Ra=10^{10}\). The vertical dashed lines indicate the number of modes required to capture \(99\%\) of the total fluctuating kinetic energy. 
  (\textit{c}) The required number of modes as a function of \(Ra\).}
\label{fig:pod_energy}
\end{figure}

At moderate \(Ra=10^{8}\) (see figure~\ref{fig:pod_energy}\textit{a}), only a few modes capture most of the energy, consistent with the presence of coherent multiple-roll LSCs. The edges of adjacent rolls, together with the plume-ejection and plume-impingement zones, act as transport barriers. Crossing these barriers requires additional propulsive acceleration, which is consistent with the sharp onset of success once \(\mathcal{A}_{\max}\) exceeds a threshold value. At higher \(Ra=10^{10}\) (see figure~\ref{fig:pod_energy}\textit{b}), the energy is distributed over more modes and the coherence weakens. The large-scale barriers become more fragmented and drift in time, forming short-lived pathways. As a result, \(S(\mathcal{A}_{\max})\) increases more gradually, and its onset shifts to higher \(\mathcal{A}_{\max}\). Figure~\ref{fig:pod_energy}(\textit{c}) further summarises the number of modes required to capture \(99\%\) of the total energy as a function of \(Ra\). These trends suggest that the success-rate curves evolve from step-like to smoother as the flow changes from a low-order, barrier-dominated regime to a more intermittent regime at higher \(Ra\).

\begin{figure}
    \centering
    \includegraphics[width=0.5\textwidth]{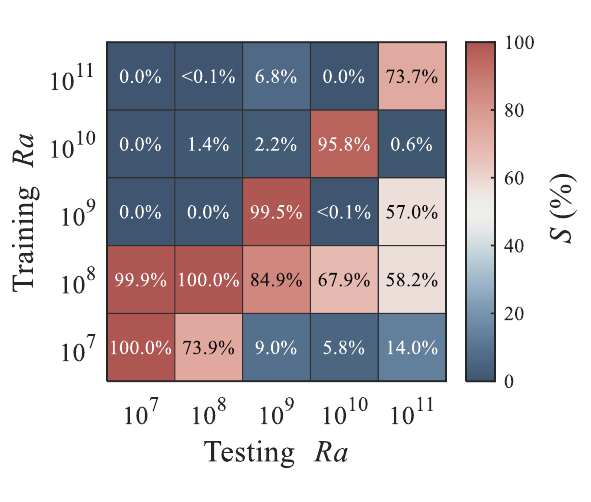}
  \caption{Generalisability of the learned policies across Rayleigh numbers. The heatmap shows the success rate \(S\) of policies trained at a given \(Ra\) (rows) and evaluated at different testing \(Ra\) (columns). All cases are evaluated at \(\mathcal{A}_{\max}=5.0\).}
    \label{fig:generalizability}
\end{figure}

To assess the robustness and generalisability of the learned strategies across different turbulence intensities, we conducted a cross-evaluation study. Policies trained at one Rayleigh number were then deployed in environments at other Rayleigh numbers. All evaluations were performed with \(\mathcal{A}_{\max}=5.0\). The resulting success rates are shown in the cross-evaluation heatmap in figure~\ref{fig:generalizability}, where the diagonal elements represent the baseline success rates when training and testing are performed in the same flow environment. Deploying a policy across full orders of magnitude in \(Ra\) represents a stringent regime shift. When evaluated under more moderate environmental variations closer to the training domain, the learned policies transfer well without any retraining. For example, a policy trained at \(Ra=10^8\) achieves a \(100.0\%\) success rate when deployed at \(Ra=1.5\times 10^8\), and a policy trained at \(Ra=10^9\) maintains a \(99.5\%\) success rate when evaluated at \(Ra=1.5\times 10^9\).
A notable feature in figure~\ref{fig:generalizability} is the asymmetry in generalisability across large \(Ra\) gaps. Policies trained at lower \(Ra\), for example \(Ra=10^8\), transfer reasonably well to higher-\(Ra\) flows, achieving a success rate of \(67.9\%\) at \(Ra=10^{10}\). By contrast, policies trained at higher \(Ra\) show much lower success rates when deployed at lower \(Ra\); for example, training at \(Ra=10^{10}\) yields a success rate of only \(1.4\%\) at \(Ra=10^8\). Because the same hyperparameter set was used across all training cases, this asymmetry does not appear to be primarily caused by case-specific hyperparameter tuning. Rather, it is consistent with changes in the underlying flow topology. This interpretation is also consistent with recent findings in active navigation \citep{Jiao2025Gradients}, where policies trained at low Reynolds numbers generalise to high Reynolds numbers more effectively than in the reverse direction. At moderate \(Ra\), the flow is dominated by coherent LSC rolls and persistent transport barriers, so the policy tends to learn a relatively direct and thrust-intensive strategy for crossing them. When transferred to a higher-\(Ra\) environment, this barrier-crossing logic can remain effective because the corresponding barriers are more fragmented and transient. Conversely, a policy trained at high \(Ra\) learns a more opportunistic strategy, mainly exploiting transient openings and plume-assisted pathways to reduce propulsion. When placed in a moderate-\(Ra\) flow, such transient openings are less common, and the policy does not appear to have learned a sufficiently robust response for crossing the more persistent transport boundaries. This leads to a substantially lower success rate.

\subsection{Energy--time trade-offs in the reward formulation} \label{sec:trade-offs}

The reward function balances time-based rewards, represented by the term \(R V_{\mathrm{eff}}^{*}(t)\), and energy-based penalties, represented by \(-Q \|\bm{a}_{\mathrm{propel}}^{*}(t)\|\). The ratio \(R/Q\) sets this trade-off and therefore influences both the temporal efficiency and the energetic cost of the self-propelled particle. When \(R/Q\) is small, the propulsion penalty dominates; in the limit \(R/Q \to 0\), the reward reduces to \(r_t=-\|\bm{a}_{\mathrm{propel}}^{*}(t)\|\). When \(R/Q\) is large, the reward for forward progress dominates; in the limit \(R/Q \to \infty\), the reward reduces to \(r_t=V_{\mathrm{eff}}^{*}(t)\). This formulation reflects the classical compromise between energy expenditure and travel time in locomotion problems. The hydrodynamics of low-Reynolds-number swimmers naturally involves a balance between mechanical power input and translational speed \citep{Lauga2009RPP}, and similar energy--time trade-offs have been discussed for active particles operating under finite energy budgets \citep{Bechinger2016RMP}. Reinforcement-learning control frameworks likewise often use composite reward functions that encourage rapid yet energy-efficient motion, as demonstrated for numerical swimmers and flow-navigation tasks \citep{Reddy2016PNAS, Colabrese2017PRL, Verma2018PNAS}. In the present context, the ratio \(R/Q\) plays an analogous role by determining whether the agent prioritises rapid progress or economical propulsion.

\begin{figure}
\centering
\includegraphics[width=0.9\textwidth]{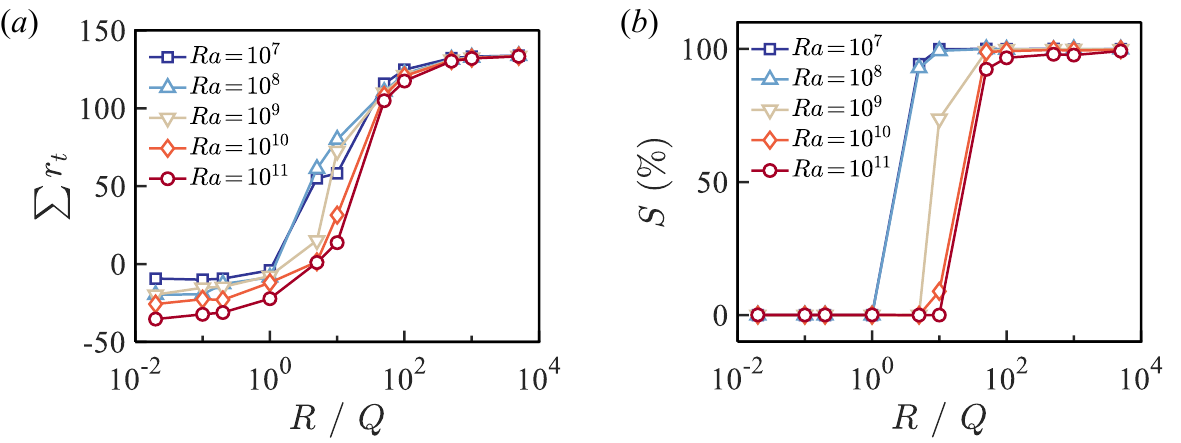}
  \caption{(\textit{a}) Average cumulative reward \(\langle \sum r_t \rangle\) and (\textit{b}) success rate \(S\) as functions of the reward-weighting ratio \(R/Q\) for different Rayleigh numbers \(Ra\).}
\label{fig:reward_ratio}
\end{figure}

Figure~\ref{fig:reward_ratio}(\textit{a}) shows how the reward weighting \(R/Q\) influences the learned policy. The averaged cumulative reward \(\langle \sum r_t \rangle\) increases with \(R/Q\). When \(R/Q \lesssim 10\), the penalty term dominates, and the return remains small or negative for all \(Ra\). As \(R/Q\) increases to \(\mathcal{O}(10)\)--\(\mathcal{O}(100)\), the return rises steeply before reaching a plateau, indicating convergence towards policies that prioritise rapid progress. The success rate in figure~\ref{fig:reward_ratio}(\textit{b}) exhibits a similar transition. Success is negligible at small \(R/Q\), increases rapidly once a threshold ratio \((R/Q)_c\) is exceeded and then saturates slightly below unity at \(Ra=10^{11}\). As \(Ra\) increases, the success rate at fixed \(R/Q\) decreases and the threshold value \((R/Q)_c\) shifts to larger values. This trend suggests that stronger emphasis on forward progress is required in more vigorous convective flows. Because the energy-based penalty discourages large control effort, a larger \(R/Q\) is required at high \(Ra\) to favour forward progress and enable barrier crossing. For example, at \(R/Q = 10\), the success rate is about \(70\%\) for \(Ra=10^{9}\), but drops to about \(10\%\) at \(Ra=10^{10}\).

\begin{figure}
\centering
\includegraphics[width=0.9\textwidth]{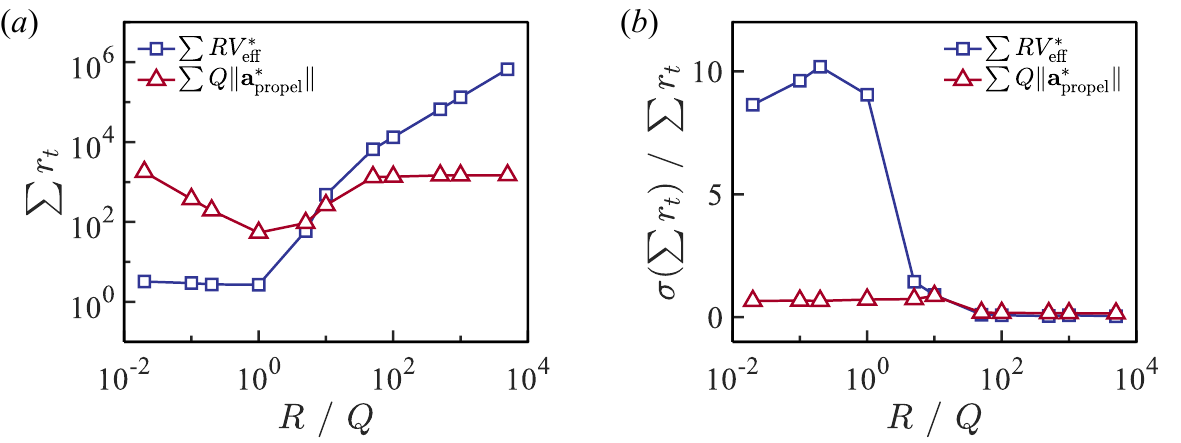}
  \caption{(\textit{a}) Average cumulative reward components and (\textit{b}) their relative variations, \(\sigma(\sum r_t)/\langle \sum r_t \rangle\). The curves show how the total reward decomposes into the time-efficiency term \(\sum R V_{\mathrm{eff}}^*\) (squares) and the actuation-penalty term \(\sum Q\|\boldsymbol{a}_{\mathrm{propel}}^*\|\) (triangles), as functions of the reward-weighting ratio \(R/Q\) at \(Ra=10^9\).}
\label{fig:reward_components}
\end{figure}

To clarify the competition between the time reward and the actuation penalty, figure~\ref{fig:reward_components} decomposes the averaged reward at \(Ra=10^{9}\) into two unnormalised cumulative contributions, namely the time-reward term \(\sum R V_{\mathrm{eff}}^*\) and the actuation-penalty magnitude \(\sum Q \|\bm{a}_{\mathrm{propel}}^*\|\). In figure~\ref{fig:reward_components}(\textit{a}), when \(R/Q \lesssim 10\), the penalty contribution dominates, consistent with the near-zero success rate observed in figure~\ref{fig:reward_ratio}(\textit{b}). Thus, a small \(R/Q\) penalises propulsion, discourages energetically costly actions and can hinder the exploitation of transport pathways in convective flows. As \(R/Q\) increases to \(\mathcal{O}(10)\)--\(\mathcal{O}(100)\), the term \(\sum R V_{\mathrm{eff}}^*\) grows rapidly, whereas \(\sum Q \|\bm{a}_{\mathrm{propel}}^*\|\) remains nearly constant. The improvement in success therefore arises mainly from an increase in the time-reward contribution rather than from a reduction in the actuation penalty.
Figure~\ref{fig:reward_components}(\textit{b}) further shows the relative variation of the total reward, \(\sigma(\sum r_t) / \langle \sum r_t \rangle\). The variation of \(\sum R V_{\mathrm{eff}}^*\) is large at small \(R/Q\) but drops sharply once \(R/Q\) exceeds \(\mathcal{O}(10)\), whereas the variation of \(\sum Q \|\bm{a}_{\mathrm{propel}}^*\|\) remains small and depends only weakly on \(R/Q\). These trends suggest that higher success rates are associated with more stable policies, because large reward variability reflects a highly stochastic policy that often fails to complete the task. Overall, this multi-objective optimisation problem exhibits a threshold weighting \((R/Q)_c\) above which training yields consistently successful policies. This threshold marks the point at which optimising forward progress begins to outweigh minimising actuation expenditure. Beyond it, the policy stabilises and both reward components vary only weakly. Accordingly, in the primary tests we use a sufficiently large value, namely \(R/Q=5000\).

\begin{figure}
\centering
\includegraphics[width=\textwidth]{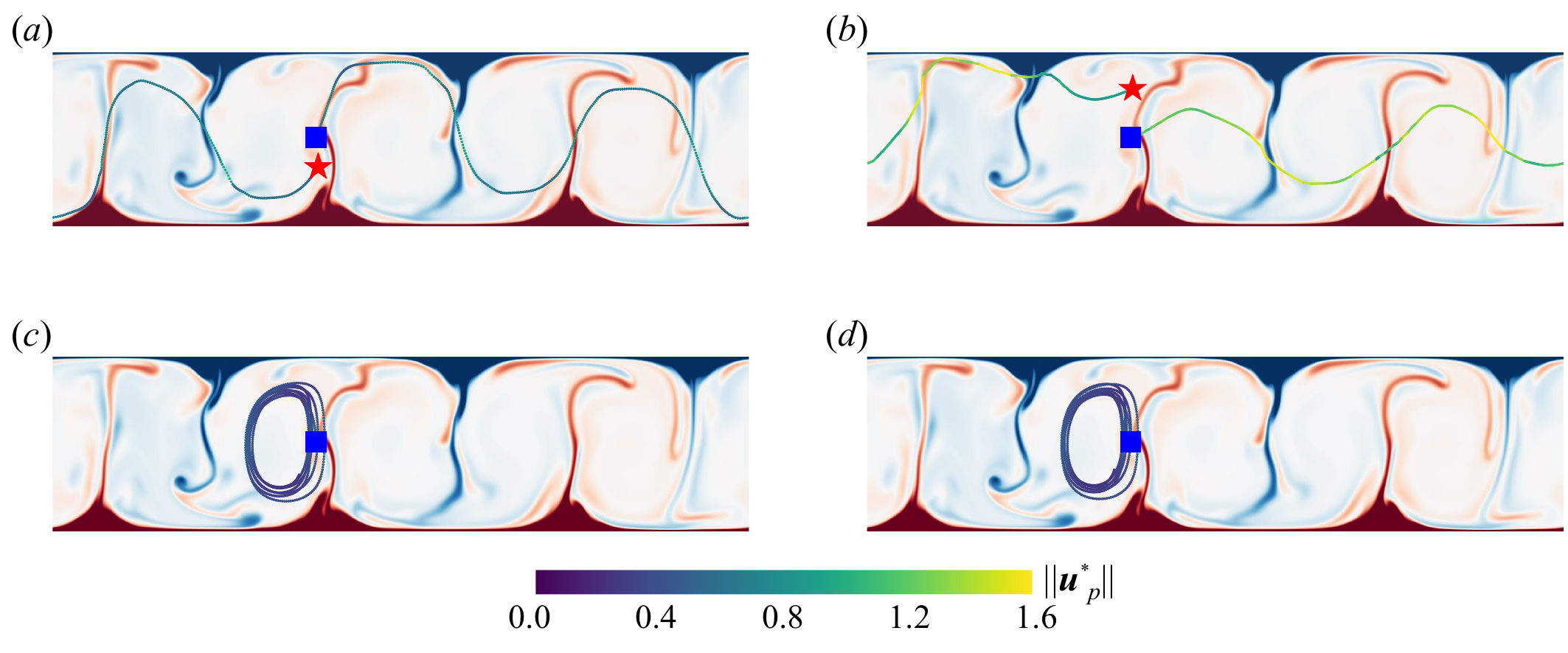}
\caption{Trajectories illustrating the limiting behaviours of the reward formulation for two actuation bounds. (\textit{a},\textit{b}) The time-reward-only limit (\(R/Q \to \infty\)) and (\textit{c},\textit{d}) the energy-penalty-only limit (\(R/Q \to 0\)), for (\textit{a},\textit{c}) \(\mathcal{A}_{\max}=2.5\) and (\textit{b},\textit{d}) \(\mathcal{A}_{\max}=10\). The background contours indicate the instantaneous dimensionless temperature \(T^*\), and the colour along each trajectory indicates the dimensionless particle speed \(\|\boldsymbol{u}_p^*\|\).}
\label{fig:reward_limits}
\end{figure}

To clarify the limiting behaviours implied by the reward decomposition, figure~\ref{fig:reward_limits} compares trajectories in the two limiting cases: time reward only (\(R/Q \to \infty\)) and energy penalty only (\(R/Q \to 0\)). In figures \ref{fig:reward_limits}(\textit{a}) and \ref{fig:reward_limits}(\textit{b}), the particle prioritises rapid progress, maintains a higher instantaneous speed (as indicated by the trajectory colouring) and follows plume-assisted pathways that cut through roll boundaries and near-wall shear layers. With \(\mathcal{A}_{\max}=2.5\), the trajectory already threads transient openings in the LSC. Increasing the bound to \(\mathcal{A}_{\max}=10\) mainly straightens the trajectory and reduces residence near stagnation regions, thereby shortening the traversal time without altering the overall strategy. While our principal quantitative analyses focus on \(\mathcal{A}_{\max}\in[0,5]\) to study flow-assisted transport, the larger bound \(\mathcal{A}_{\max}=10\) is used here only for illustration. It demonstrates the limiting behaviour when the control amplitude is sufficiently large, suggesting that the main features of these reward-driven strategies persist even when the agent can readily overpower the local flow.
Figures~\ref{fig:reward_limits}(\textit{c}) and \ref{fig:reward_limits}(\textit{d}) show the complementary condition in which only the energy penalty is applied. In this case, the RL-controlled self-propelled particle minimises its propulsive acceleration, aligns with the local circulation of the large-scale roll and is advected onto nearly closed orbits that avoid plume-ejection and plume-impingement regions. The particle repeatedly recirculates within a single convection cell, lingering near separatrices and stagnation saddles rather than using propulsive effort to cross them. Even when the actuation bound is increased from \(\mathcal{A}_{\max}=2.5\) (see figure~\ref{fig:reward_limits}\textit{c}) to \(\mathcal{A}_{\max}=10\) (see figure~\ref{fig:reward_limits}\textit{d}), the motion remains confined to the roll and the prescribed horizontal displacement is not completed within the allowed time. These examples show that, when rapid progress is not encouraged, energy-saving behaviour leads to attachment to the LSC and poor reachability, whereas time-oriented control exploits intermittent transport pathways to cross barriers. A higher \(\mathcal{A}_{\max}\) then acts mainly to accelerate crossing rather than to change the underlying navigation logic.

\subsection{Navigation mechanisms: kinematics and Lagrangian coherent structures}

\begin{figure}
\centering
\includegraphics[width=0.9\textwidth]{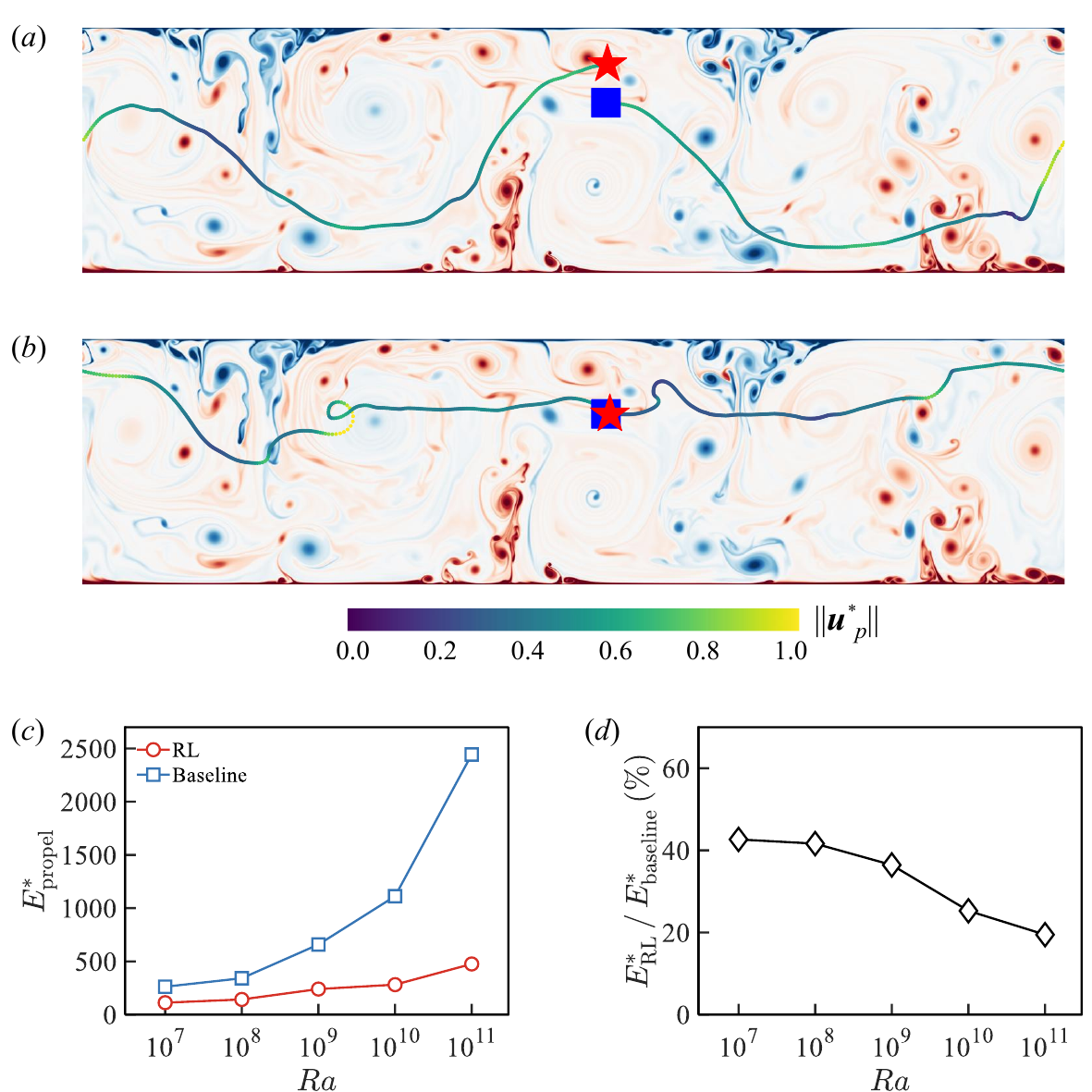}
\caption{Comparison of the learned RL policy with the constant-heading baseline for the fixed-displacement task. Trajectories obtained with (\textit{a}) the RL policy and (\textit{b}) the constant-heading baseline at \(Ra=10^{10}\) and \(\mathcal{A}_{\max}=5.0\). The background contours indicate the instantaneous dimensionless temperature \(T^*\), and the colour along each trajectory indicates the dimensionless particle speed \(\|\boldsymbol{u}_p^*\|\). As in figure~\ref{fig:navigation_trajectories}, the start marker (blue square) and goal marker (red star) have the same horizontal coordinate because of the periodic lateral boundaries; in  (\textit{b}), they overlap visually because the particle returns to its initial vertical position on completing the task. (\textit{c}) The dimensionless propulsion energy \(E_{\mathrm{propel}}^*\) versus \(Ra\) for both strategies. (\textit{d}) The relative propulsion cost \(E_{\mathrm{RL}}^*/E_{\mathrm{baseline}}^*\) as a function of \(Ra\).}
\label{fig:rl_vs_baseline}
\end{figure}

Figure~\ref{fig:rl_vs_baseline} compares the learned RL policy with a constant-heading baseline for the fixed-displacement task at \(Ra=10^{10}\) and \(\mathcal{A}_{\max}=5.0\). Details of the baseline configuration are given in \S~\ref{sec:appNaive}. The RL agent follows plume-assisted pathways near roll edges and avoids recirculation zones (see figure~\ref{fig:rl_vs_baseline}\textit{a}), whereas the baseline agent maintains a constant heading towards the target and repeatedly encounters adverse cross-stream currents (see figure~\ref{fig:rl_vs_baseline}\textit{b}). To aid physical interpretation, side-by-side visualisations of these distinct navigation strategies at moderate and high Rayleigh numbers are provided in supplementary movie~1 (\(Ra=10^{8}\)) and supplementary movie~2 (\(Ra=10^{10}\)) available at \url{https://doi.org/10.1017/jfm.2026.11802}.

These trajectory differences are reflected in the propulsion energetics (see figure~\ref{fig:rl_vs_baseline}\textit{c}). For a fair comparison, we define a reference actuation level \(\mathcal{A}_{\max}\) for each \(Ra\) as the value that yields a success rate of \(80\%\) in the convective flow. A sensitivity analysis, not shown for brevity, indicates that this energetic advantage remains evident under a stricter success criterion, for example \(90\%\). The propulsion energy \(E^*_{\mathrm{propel}}\) of the RL policy increases only slightly with \(Ra\), whereas that of the constant-heading baseline rises sharply. Consequently, the ratio \(E^*_{\mathrm{RL}}/E^*_{\mathrm{baseline}}\) decreases from about \(45\%\) at \(Ra=10^{7}\) to \(17\%\) at \(Ra=10^{11}\) (see figure~\ref{fig:rl_vs_baseline}\textit{d}), suggesting that the relative advantage of the learned policy becomes more pronounced with increasing \(Ra\). This occurs because the learned policy preferentially aligns the particle with favourable flow structures, allowing part of the mechanical work to be supplied by the background flow through drag. By contrast, the constant-heading baseline encounters strong counterflows more frequently as \(Ra\) increases, so the opposing streamwise velocity and enhanced shear lead to substantially higher propulsion energy.

\begin{figure}
\centering
\includegraphics[width=0.9\textwidth]{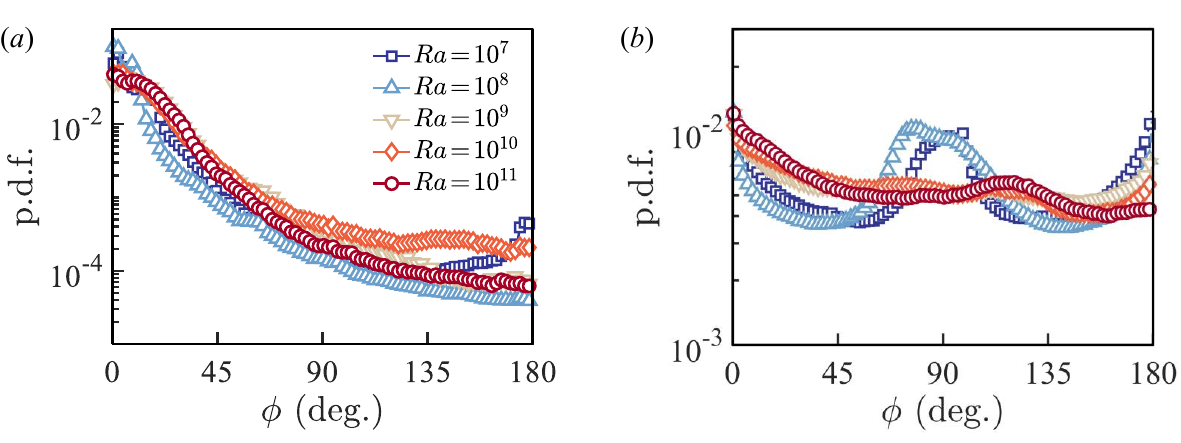}
\caption{The p.d.f.s of the instantaneous alignment angle \(\phi\) between the particle's propulsive vector and the local fluid-velocity vector. The data are computed over all successful trajectories obtained with (\textit{a}) the learned RL policy and (\textit{b}) the constant-heading baseline for different Rayleigh numbers \(Ra\).}
\label{fig:alignment_pdf}
\end{figure}

To further compare propulsive behaviour during navigation, we examine the probability density functions (p.d.f.s) of the instantaneous alignment angle \(\phi\) between the particle's propulsive vector and the local fluid-velocity vector, computed over all successful trials. A value of \(\phi=0^{\circ}\) indicates alignment with the carrier flow, whereas \(\phi=180^{\circ}\) indicates opposition. For the learned strategy, the p.d.f. exhibits a pronounced peak near \(0^{\circ}\) and then decreases monotonically (see figure~\ref{fig:alignment_pdf}\textit{a}). This pattern suggests that the RL agent tends to align its propulsion direction with the local current and thereby exploits the carrier flow to progress towards the target. The tail broadens slightly as the Rayleigh number increases, consistent with increasingly intermittent reorientation of the local flow. The constant-heading baseline yields a comparatively flat distribution with a small secondary peak near \(90^{\circ}\) (see figure~\ref{fig:alignment_pdf}\textit{b}), reflecting frequent cross-stream events. The baseline agent therefore often accelerates nearly orthogonally to the carrier flow rather than following it, which leads to higher control effort and is consistent with the larger propulsive-energy expenditure reported in figure~\ref{fig:rl_vs_baseline}.

\begin{figure}
    \centering
    \includegraphics[width=0.9\textwidth]{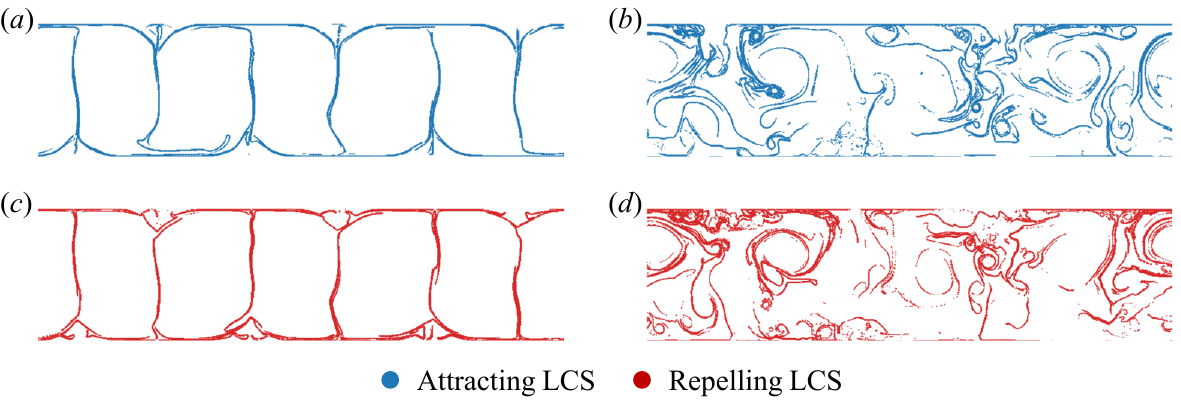}
\caption{Instantaneous snapshots of the LCS at \(t^*=9.00\) for (\textit{a},\textit{c}) \(Ra=10^8\) and (\textit{b},\textit{d}) \(Ra=10^{10}\).  (\textit{a},\textit{b}) The attracting LCS (blue regions).  (\textit{c},\textit{d}) The repelling LCS (red regions).}
    \label{fig:instant_lcs}
\end{figure}

To better understand the physical origin of this energetic advantage, we move from an Eulerian kinematic description to a finite-time Lagrangian view of transport barriers and pathways. In unsteady flows, such structures can be characterised using LCS, which are identified here as ridges of the finite-time Lyapunov exponent (FTLE) fields \citep{Krishna2022PRSA,krishna2023finite}. To relate the learned trajectories to these structures, we compute both forward-time and backward-time FTLE fields for the convective flow. For visualisation, the LCS fields are extracted by thresholding the FTLE fields: regions exceeding \(70\%\) of the maximum forward- and backward-time FTLE values are classified as repelling and attracting LCS, respectively.
As shown in figure~\ref{fig:instant_lcs}, at moderate Rayleigh number (\(Ra=10^8\)), the repelling LCS, which act as finite-time transport barriers, and the attracting LCS, which indicate preferred transport pathways, show substantial spatial overlap. Together they outline closed or nearly closed boundaries around the LSC cells. This topological organisation is consistent with the trapping observed for the energy-penalty-only policy: because this policy applies little propulsive thrust, it tends to remain within the same circulation cell rather than crossing the repelling FTLE ridges. At higher Rayleigh number (\(Ra=10^{10}\)), this overlap weakens, and the repelling LCS no longer form comparably closed structures. The associated weakening and fragmentation of the barrier network provide more intermittent pathways through which the agent can traverse the domain.

\begin{figure}
    \centering
    \includegraphics[width=0.9\textwidth]{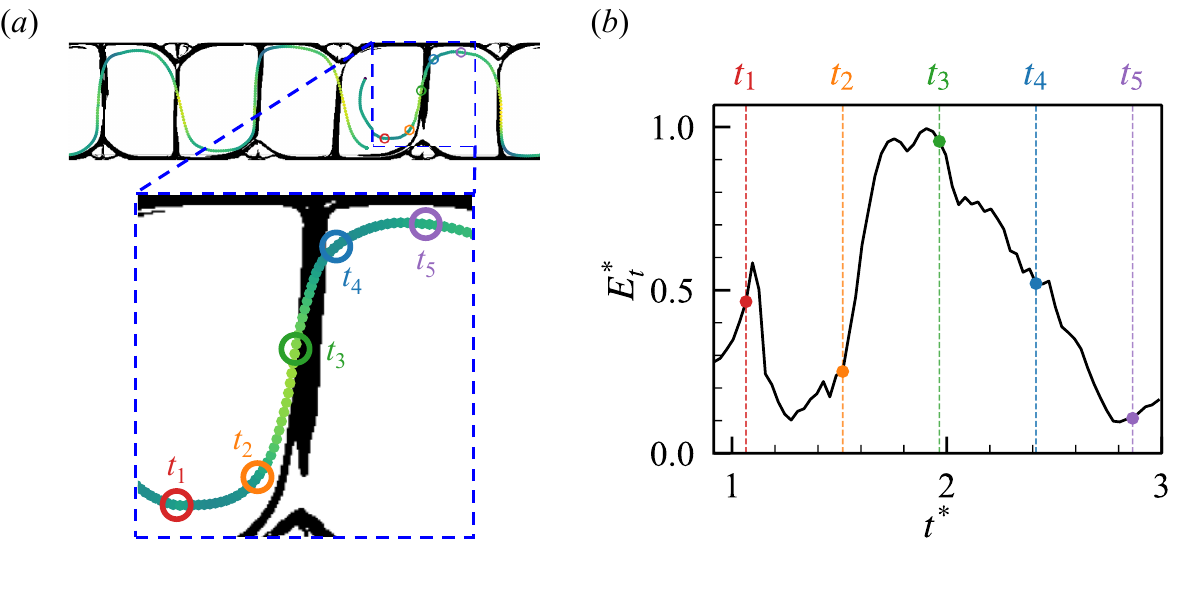}
\caption{(\textit{a}) Representative trajectory of the RL agent at \(Ra=10^8\) with \(\mathcal{A}_{\max}=1.25\), corresponding to the learned policy used for this regime. The upper panel shows the repelling LCS over the full domain, whereas the lower panel provides a magnified view of the dashed region and highlights a barrier-crossing event. Coloured markers indicate five representative instants, \(t_1\)--\(t_5\), along the trajectory. (\textit{b}) Corresponding time history of the dimensionless instantaneous propulsion cost \(E_t^*\) as a function of dimensionless time \(t^*\).}
    \label{fig:lcs_crossing_ra1e8}
\end{figure}

\begin{figure}
    \centering
    \includegraphics[width=0.9\textwidth]{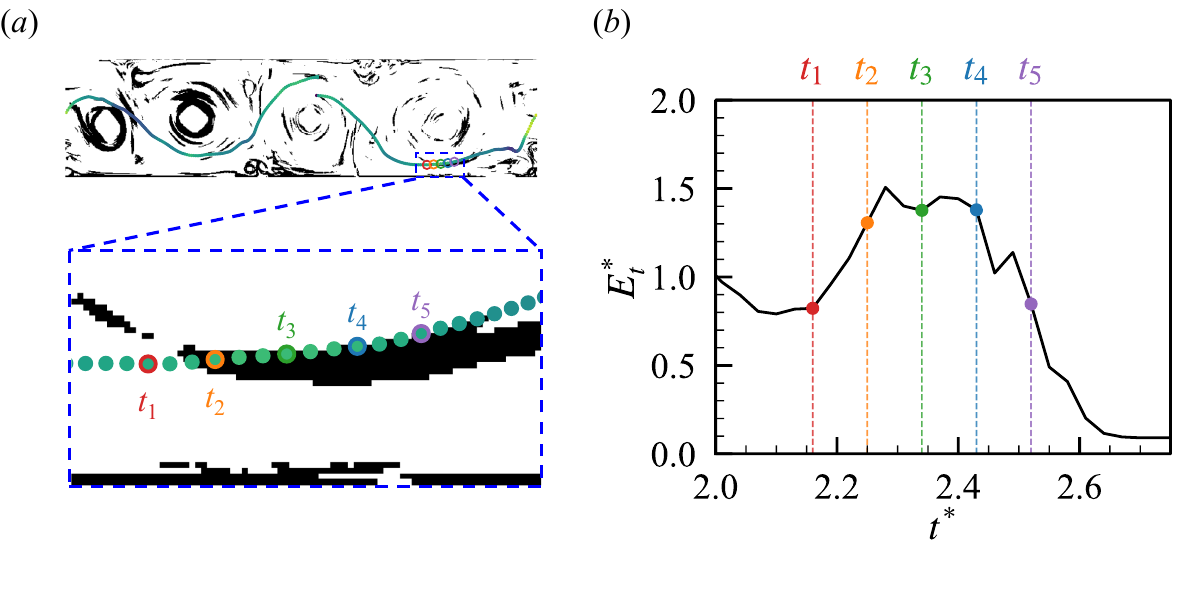}
\caption{(\textit{a}) Representative trajectory of the RL agent at \(Ra=10^{10}\) with \(\mathcal{A}_{\max}=4.0\), corresponding to the learned policy used for this regime. The upper panel shows the repelling LCS over the full domain, whereas the lower panel provides a magnified view of the dashed region and highlights a typical event in which the trajectory crosses a repelling LCS. Coloured markers indicate five representative instants, \(t_1\)--\(t_5\), along the trajectory. (\textit{b}) Corresponding time history of the dimensionless instantaneous propulsion cost \(E_t^*\) as a function of dimensionless time \(t^*\).}
    \label{fig:lcs_crossing_ra1e10}
\end{figure}

Furthermore, because crossing an LCS is a dynamic and continuous process, comparing a trajectory with a single instantaneous FTLE snapshot is insufficient. To analyse the barrier-crossing events, we therefore compute a time-averaged FTLE field over the corresponding crossing interval, for example from \(t_1\) to \(t_5\). Applying a threshold of \(60\%\) of the maximum value to this time-averaged field is used to extract the persistent LCS.
Figures~\ref{fig:lcs_crossing_ra1e8} and~\ref{fig:lcs_crossing_ra1e10} show a representative learned particle trajectory overlaid on the corresponding time-averaged FTLE ridge during a barrier-crossing event. The corresponding instantaneous propulsion cost \(E_t^*\) is plotted alongside. As the particle approaches and actively crosses the repelling FTLE ridge, i.e. the barrier, between \(t_2\) and \(t_4\), the propulsion cost exhibits a pronounced local increase. This suggests that the agent applies large propulsive effort near this topological bottleneck. Immediately after crossing the barrier, the particle aligns with the attracting LCS, i.e. the transport pathway, and its propulsion cost decreases as it follows favourable downstream currents.
These results suggest that the RL policy exploits aspects of the underlying Lagrangian topology of the turbulent convective flow. We acknowledge, however, that the representative trajectory overlays and time-averaged FTLE fields, although supportive of the barrier-crossing interpretation, do not yet constitute a full event-level statistical census of all crossings. Such a quantitative analysis would require automated identification of individual trajectory intersections with repelling FTLE ridges, followed by correlation with local propulsion-cost peaks. This is non-trivial in the high-\(Ra\) regimes, because the LCS geometry is highly transient and the topological extraction can be sensitive to the chosen finite-time integration window. We therefore regard a rigorous automated event-level analysis of LCS crossings as an important direction for future research.
Furthermore, although FTLE analysis provides a mathematically well-defined description of finite-time transport barriers, its computation requires non-local, time-resolved flow information over a finite integration window. By contrast, the RL agent navigates these dynamic barriers using only instantaneous, locally measurable cues. To understand how the policy achieves this, we next translate these global Lagrangian structures into local Eulerian observables.

\subsection{Distilling an interpretable heuristic strategy from Eulerian flow topology}
\label{sec:heuristic}

Although the RL policy achieves lower propulsion cost than the constant-heading baseline, the physical logic underlying this black-box optimisation still warrants clearer interpretation. To elucidate how the agent exploits the observed flow information, we analyse the particle behaviour relative to the local flow structures, focusing on representative cases at \(Ra=10^8\) and \(Ra=10^{10}\) under their respective reference actuation bounds.

\begin{figure}
    \centering
    \includegraphics[width=\textwidth]{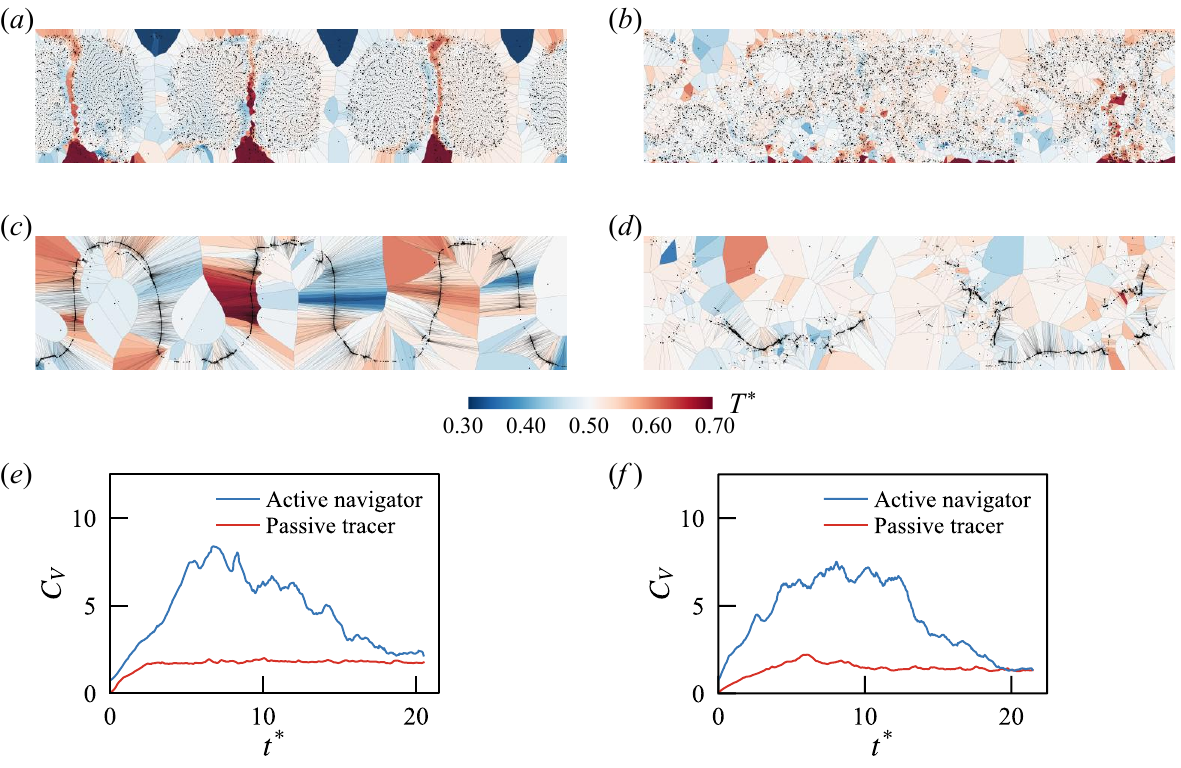}   
\caption{Instantaneous Voronoi tessellations of the distributions of (\textit{a},\textit{b}) passive tracers and (\textit{c},\textit{d}) active navigators at \(t^*=9.00\). (\textit{e},\textit{f}) The corresponding temporal evolution of the clustering index \(C_V\) for the reference actuation conditions. The Voronoi cells are coloured by the instantaneous background temperature field \(T^*\). (\textit{a},\textit{c},\textit{e}) Rayleigh number \(Ra=10^8\) with \(\mathcal{A}_{\max}=1.25\) and (\textit{b},\textit{d},\textit{f})  \(Ra=10^{10}\) with \(\mathcal{A}_{\max}=4.0\).}
    \label{fig:voronoi_clustering}
\end{figure}

Our analysis shows that the learned policy exhibits a consistent behavioural pattern. To highlight the agent's spatial preferences, we compare the distribution of active navigators with that of passive tracers released under identical conditions. The instantaneous Voronoi tessellations in figure~\ref{fig:voronoi_clustering}(\textit{a}--\textit{d}) show that the passive tracers (figures~\ref{fig:voronoi_clustering}\textit{a} and \ref{fig:voronoi_clustering}\textit{b}) are broadly dispersed by the background turbulence and often become trapped within recirculating vortex cores. By contrast, the active navigators (figures~\ref{fig:voronoi_clustering}\textit{c} and \ref{fig:voronoi_clustering}\textit{d}) preferentially avoid these cores and instead accumulate near the outer regions of vortical structures and along roll edges, where pathways for horizontal traversal are more favourable. To quantify this preferential clustering, we introduce a clustering index \(C_V\), defined as the normalised standard deviation of the Voronoi-cell areas:

\begin{equation}
C_V = \sqrt{\frac{1}{M} \sum_{i=1}^{M} \left( \frac{A_i}{\bar{A}} - 1 \right)^2},
\end{equation}
where \(A_i\) denotes the area of the \(i\)th Voronoi cell, \(\bar{A}\) is the mean Voronoi-cell area and \(M\) is the total number of particles. As shown in figures~\ref{fig:voronoi_clustering}(\textit{e}) and \ref{fig:voronoi_clustering}(\textit{f}), \(C_V\) for the active navigators remains consistently higher than for the passive tracers throughout the migration process. This larger variance in cell areas suggests that the learned policy steers particles away from broad, relatively featureless flow regions, leaving a few large Voronoi cells, while concentrating them within narrow and dynamically favourable pathways, thereby producing dense clusters of small cells. In this way, rather than being passively mixed by the flow, the active navigators exploit the spatial heterogeneity of the turbulence to their advantage.

\begin{figure}
    \centering
    \includegraphics[width=0.9\textwidth]{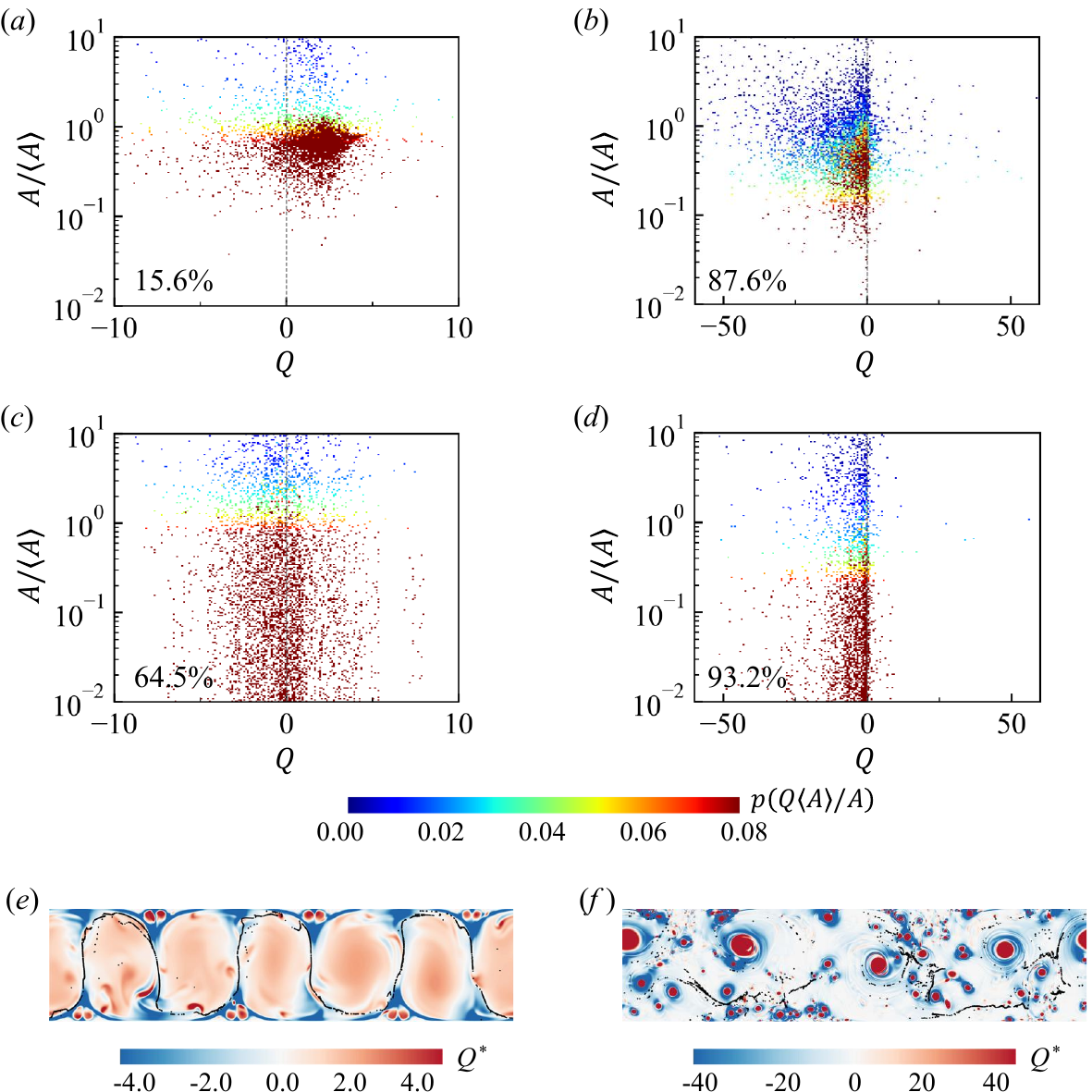}
\caption{Joint probability distributions of the Voronoi-cell area \(A\) and the local \(Q\)-value at \(t^*=9.00\) for the reference actuation conditions. 
Passive tracers at (\textit{a}) \(Ra=10^8\) and (\textit{b}) \(Ra=10^{10}\);
(\textit{c},\textit{d}) active navigators at the corresponding Rayleigh numbers. The colour scale in (\textit{a}--\textit{d}) denotes the joint probability density \(p(Q, A/\langle A \rangle)\), where \(\langle A \rangle\) is the mean Voronoi-cell area.  The corresponding instantaneous particle distributions at \(t^*=9.00\), overlaid on the instantaneous \(Q^*\) field for (\textit{e}) \(Ra=10^8\) and (\textit{f}) \(Ra=10^{10}\).}
    \label{fig:voronoi_q}
\end{figure}

To characterise the nature of these favourable pathways, we examine the coupling between particle distribution and local flow topology by plotting the joint distribution of the Voronoi-cell area \(A\) and the local \(Q\)-value, as shown in figure~\ref{fig:voronoi_q}. The \(Q\)-criterion distinguishes rotation-dominated regions (vortex cores, \(Q>0\)) from strain-dominated regions (shear layers and transport pathways, \(Q<0\)). The scatter plots show that, at a moderate Rayleigh number (\(Ra=10^8\)), passive tracers are mainly trapped in rotation-dominated cores, with only \(15.6\%\) located in the \(Q<0\) regime (see figure~\ref{fig:voronoi_q}\textit{a}). By contrast, active navigators are redistributed away from these traps, shifting a large fraction of their distribution (\(64.5\%\)) into the strain-dominated \(Q<0\) regions (see figure~\ref{fig:voronoi_q}\textit{c}). Furthermore, the high-probability-density region (red) simultaneously shifts towards smaller Voronoi-cell areas, \(A/\langle A\rangle < 1\), suggesting that active navigators not only avoid vortical regions but also cluster within narrow and favourable pathways.
Even at \(Ra=10^{10}\), where the highly turbulent background flow already scatters most passive tracers into \(Q<0\) regions (\(87.6\%\); see figure~\ref{fig:voronoi_q}\textit{b}), the active navigators further increase this fraction to \(93.2\%\) (see figure~\ref{fig:voronoi_q}\textit{d}). More importantly, comparison of figures~\ref{fig:voronoi_q}(\textit{b}) and~\ref{fig:voronoi_q}(\textit{d}) shows that the RL policy transforms a dispersed passive distribution into a more concentrated cluster at low \(A/\langle A\rangle\), suggesting pronounced spatial agglomeration within strain-dominated regions. By actively avoiding \(Q>0\) regions (see figures~\ref{fig:voronoi_q}\textit{e} and \ref{fig:voronoi_q}\textit{f}), the active navigators reduce closed-orbit recirculation and remain in the strain-dominated outer layers, which are favourable for sustained downstream migration. This interpretation is consistent with our trajectory analysis, which showed that the learned strategy tends to follow plume-assisted pathways and remain clustered near fragmented transport barriers.

To connect the Eulerian and Lagrangian interpretations, it is useful to clarify how the instantaneous \(Q\)-criterion relates to the FTLE fields. These two diagnostics are complementary but not equivalent. The \(Q\)-criterion provides a local and instantaneous Eulerian classification of the flow topology, with \(Q>0\) corresponding to rotation-dominated vortex cores and \(Q<0\) to strain-dominated regions. By contrast, FTLE ridges identify finite-time LCS that organise material transport over a finite integration window. In the present flow, the \(Q>0\) vortex cores are often located inside regions enclosed or constrained by repelling LCS, whereas the surrounding \(Q<0\) regions are more closely associated with stretching and inter-roll transport. This correspondence provides a physical link between the two analyses: by avoiding \(Q>0\) regions, the active navigators reduce the likelihood of remaining trapped in recirculating vortex cores; by staying closer to strain-dominated regions, they are more likely to access transport pathways identified in the LCS analysis. Importantly, computing FTLE fields requires non-local, time-resolved velocity information over a finite integration window, which is not directly available to an autonomous agent making real-time decisions. The agent instead relies on instantaneous, locally measurable Eulerian cues, including the velocity-gradient information from which \(Q\)-like topology can be inferred. The consistency among the learned trajectories, the FTLE/LCS fields and the Voronoi statistics therefore suggests that local Eulerian flow topology may serve as a practical proxy for aspects of the finite-time Lagrangian transport structure.

Motivated by these physical observations, we distil a simple heuristic strategy from the behaviour of the learned policy. The key idea is that the particle switches between two modes: a drift mode, in which horizontal propulsion is switched off to avoid inefficient thrust expenditure when the particle is located in a locally adverse but weakly constrained flow region; and a thrust mode, in which stronger horizontal propulsion is applied to escape confinement, cross barriers or maintain progress through unfavourable flow structures. For the horizontal-migration task, we formalise this heuristic strategy using two threshold parameters. Because the absolute magnitudes of the fluid velocity and velocity gradients vary substantially across the Rayleigh numbers considered here, using dimensional thresholds would require case-dependent calibration. For example, the magnitude of the \(Q\)-criterion differs by about one order of magnitude between \(Ra=10^8\) and \(Ra=10^{10}\), as shown in figures~\ref{fig:voronoi_q}(\textit{e}) and~\ref{fig:voronoi_q}(\textit{f}). To avoid threshold tuning for each \(Ra\), we define the trigger conditions using locally normalised, dimensionless variables, namely the normalised \(Q\)-criterion \(\widetilde{Q}=Q/\|\nabla \boldsymbol{u}\|_F^2\) and the normalised vertical velocity \(\widetilde{u}_y=u_y/\|\boldsymbol{u}\|_2\). These bounded quantities, \(\widetilde{Q}\in[-0.5,0.5]\) and \(\widetilde{u}_y\in[-1,1]\), characterise the relative dominance of local rotation and vertical motion rather than their absolute magnitudes. A single fixed set of thresholds is used for all Rayleigh numbers considered in this study:

\begin{equation}
\widetilde{Q}_{\text{th}}=0.3,
\qquad
\widetilde{u}_{y,\text{th}}=0.2 .
\end{equation}

The active control acceleration \(\boldsymbol{a}_{\text{propel}}=(a_{\text{propel},x},a_{\text{propel},y})\) is then defined as

\begin{equation}
a_{{\text{propel}},x}=
\begin{cases}
a_{\text{drift}}, & \text{if } \widetilde{Q}<\widetilde{Q}_{\text{th}},\ |\widetilde{u}_{y}|<\widetilde{u}_{y,\text{th}},\ \text{and } u_x<0, \\
a_{\text{thrust}}, & \text{otherwise},
\end{cases}
\end{equation}
\begin{equation}
a_{{\text{propel}},y}=\frac{\rho_p-\rho_f}{\rho_p}g,
\end{equation}
where \(a_{\text{drift}}=0\), and the thrust-mode acceleration is

\begin{equation}
a_{\text{thrust}}=
\sqrt{\|\boldsymbol{a}_{\text{propel}}\|^{2}_{\max}-a_{{\text{propel}},y}^2}.
\end{equation}

The drift condition \(\widetilde{Q}<\widetilde{Q}_{\text{th}}\), \(|\widetilde{u}_{y}|<\widetilde{u}_{y,\text{th}}\) and \(u_x<0\) corresponds to a state in which the particle is outside strongly rotation-dominated vortex cores, does not encounter strong vertical motion, but is embedded in a locally adverse horizontal flow. In this situation, applying strong positive horizontal thrust would mainly oppose the local current and lead to inefficient energy expenditure. The heuristic therefore sets \(a_{\text{drift}}=0\), allowing the particle to drift until the local flow carries it out of the adverse region. In all other cases, including proximity to rotation-dominated regions, strong vertical motion or situations requiring active barrier crossing, the particle switches to the thrust mode and applies constant horizontal propulsion. The thresholds used in this heuristic strategy were calibrated only in the lower-\(Ra\), more coherent cases, namely \(Ra=10^7\) and \(Ra=10^8\), through parameter sweeps of \(\widetilde{Q}_{\text{th}}\) and \(\widetilde{u}_{y,\text{th}}\). The same values were then applied without further tuning to the higher-\(Ra\) cases, \(Ra=10^9\), \(10^{10}\) and \(10^{11}\). Thus, the high-\(Ra\) performance can be interpreted as a transfer test of the heuristic strategy outside its calibration regime.

\begin{figure}
    \centering
    \includegraphics[width=\textwidth]{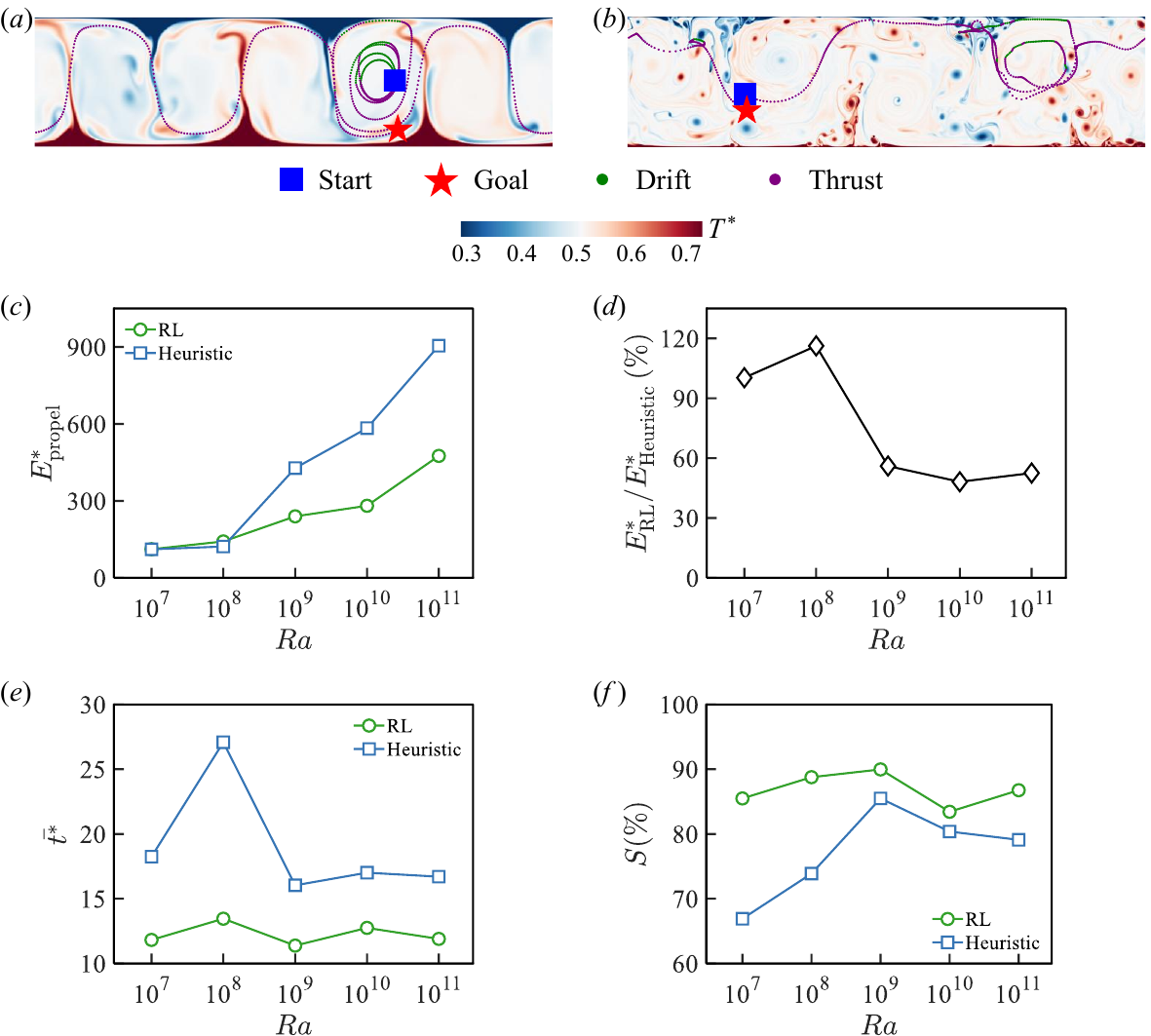}    
\caption{Comparison of representative trajectories and performance between the distilled heuristic strategy and the learned RL policy. 
(\textit{a},\textit{b}) Representative trajectories of the heuristic strategy at the reference actuation bounds: (\textit{a}) \(Ra=10^8\) with \(\mathcal{A}_{\max}=1.25\) and (\textit{b}) \(Ra=10^{10}\) with \(\mathcal{A}_{\max}=4.0\). The background colour indicates the instantaneous temperature field \(T^*\). The blue square and red star mark the start and goal locations, respectively. Green solid lines denote the drift mode, whereas magenta dotted lines denote the thrust mode. 
(\textit{c}) The propulsion energy \(E_{\mathrm{propel}}^*\) as a function of \(Ra\) for both the RL policy and the heuristic strategy. 
(\textit{d}) The corresponding energy ratio \(E_{\mathrm{RL}}^*/E_{\mathrm{Heuristic}}^*\). 
(\textit{e}) The mean completion time \(\bar{t}^*\) and (\textit{f}) the success rate \(S\) as functions of \(Ra\). The heuristic strategy uses a single fixed set of normalised thresholds for all Rayleigh numbers considered here, with \(\widetilde{Q}_{\text{th}}=0.3\) and \(\widetilde{u}_{y,\text{th}}=0.2\).}
    \label{fig:heuristic_vs_rl}
\end{figure}

Representative trajectories obtained from this distilled heuristic strategy are shown in figures~\ref{fig:heuristic_vs_rl}(\textit{a}) and \ref{fig:heuristic_vs_rl}(\textit{b}), illustrating the switching between the green drift mode and the magenta thrust mode. To contextualise the performance of the learned RL policy, we compare its energetics with those of this rule-based heuristic strategy, as shown in figures~\ref{fig:heuristic_vs_rl}(\textit{c}) and \ref{fig:heuristic_vs_rl}(\textit{d}). At high Rayleigh numbers (\(Ra \geq 10^9\)), the learned RL policy requires less propulsion energy than the heuristic baseline, suggesting that the neural-network policy can exploit highly intermittent and spatiotemporally chaotic flow structures more effectively. Interestingly, at moderate Rayleigh number (\(Ra=10^8\)), the heuristic strategy occasionally consumes less propulsion energy than the RL policy. However, this energetic advantage comes at the cost of longer completion times (see figure~\ref{fig:heuristic_vs_rl}\textit{e}). The rule-based heuristic agent can be overly conservative, switching off propulsion and becoming temporarily trapped within large, coherent LSC vortices. Because the RL reward function is calibrated to prioritise time efficiency and barrier crossing, the RL policy chooses to expend slightly more energy to escape these vortex traps more actively, thereby completing the task much faster.
Furthermore, as shown in figure~\ref{fig:heuristic_vs_rl}(\textit{f}), although this heuristic strategy is much simpler than the learned RL policy and relies on only three interpretable criteria, it reproduces the main transport logic of the learned policy. It should be noted that the distilled heuristic is intended primarily as an interpretability tool and a simple rule-based controller derived from the flow-topology analysis of the learned policy, rather than as a completely independent baseline or a universally optimal controller. The comparison shows that key features of the learned behaviour can be related to local Eulerian observables such as the \(Q\)-criterion and the vertical velocity. The fact that the heuristic retains high success rates at higher Rayleigh numbers suggests that the normalised local topology and velocity cues capture features of the navigation problem that remain useful across the range of flow regimes considered here. Thus, the RL policy is not merely an opaque numerical optimiser, but can provide a route for extracting interpretable navigation rules with reasonable transferability across the present \(Ra\) range.

\section{Conclusion}\label{sec:Conclusion}

In this work, we investigated the navigation of a self-propelled inertial particle in two-dimensional RB convection over the range \(10^7 \le Ra \le 10^{11}\) at \(Pr=0.71\) and \(\Gamma=4\). An RL controller selected the propulsive acceleration, subject to an upper bound, to complete a fixed-displacement task. We evaluated macroscopic navigation performance in terms of reachability, completion time and propulsion energy, and interpreted the underlying physical mechanisms using Eulerian flow topology and LCS.

Our main findings on macroscopic performance are as follows. The success rate increases monotonically with the actuation limit \(\mathcal{A}_{\max}\). At moderate \(Ra\), this increase is abrupt, whereas at higher \(Ra\) the transition becomes more gradual and shifts to larger \(\mathcal{A}_{\max}\). Although higher-\(Ra\) flows require stronger actuation to achieve reachability and lead to longer completion times, the energetic cost of traversal decreases once the task becomes feasible. Proper orthogonal decomposition indicates that these trends are associated with reorganisation of the carrier flow. At moderate \(Ra\), kinetic energy is concentrated in a few modes, and the multi-roll LSC creates persistent transport barriers that require a finite thrust surplus to cross. At higher \(Ra\), energy is distributed across many modes, causing the barriers to fragment and intermittent plume-assisted pathways to emerge.

We further identified the kinematic and topological strategies that enable flow-assisted navigation. The learned policy balances progress and energy expenditure by aligning its thrust with favourable local currents. Lagrangian analysis suggests that the agent modulates its propulsion to cross repelling LCS barriers and then follow attracting transport pathways. By mapping this behaviour onto the local Eulerian flow fields via Voronoi tessellation and the \(Q\)-criterion, we found that the agent preferentially avoids rotation-dominated vortex cores (\(Q>0\)) and exploits strain-dominated regions (\(Q<0\)). Based on these insights, we distilled an interpretable, physics-based heuristic strategy that reproduces the main barrier-crossing logic of the learned policy, retaining good navigability using only instantaneous and locally measurable cues.

It is important to note that this flow-assisted navigation strategy is tied to the heavy-particle, low-Stokes-number regime considered here, with \(St\sim\mathcal{O}(10^{-3})\). In this regime, the particle rapidly equilibrates with the local flow, and unsteady hydrodynamic effects such as added mass and Basset history forces remain small. For particles with larger inertia, for example \(St\sim\mathcal{O}(1)\), finite-size effects, added mass and history forces may become important. The particle would then respond less directly to the local flow, and the navigation logic may shift from the present flow-assisted, topology-sensitive behaviour towards a more inertia-dominated crossing strategy. Similarly, when rotational dynamics is important, additional orientational control would be required to counteract flow-induced torques, which could modify the energy--time trade-off.

Although the present framework examines flow-assisted translational navigation, several idealisations and scope limitations should be noted. First, the learned policy is direction-specific by construction. Because the convective flow structures, including updrafts, downdrafts and large-scale rolls, as well as the domain boundaries, are anisotropic, a policy trained only for horizontal traversal is not expected to retain the same performance when deployed in arbitrary directions. Second, the use of absolute position to track fixed-displacement progress limits the expected generalisability of the policy across modified domain geometries. The spatiotemporal variability of the flow reduces the possibility of simply memorising static flow features, but changes in the geometric configuration may still affect performance. For example, in domains with different aspect ratios, the local gradient-based navigation logic may remain partly useful, whereas the modified organisation of the LSC could change the preferred transport pathways and reduce the overall success rate. In non-periodic horizontal domains with solid lateral walls, additional effects such as lateral boundary layers and corner recirculation would arise; a policy trained only in a periodic domain may not have learned the behaviours needed to navigate such boundaries.

The two-dimensionality of the present set-up also frames the scope of the transport-barrier interpretation. In three-dimensional RB convection, the LSC may exhibit spanwise deformation, torsion and more complex plume dynamics. The corresponding boundary of LCS would also change from one-dimensional repelling curves to two-dimensional material surfaces. As a result, an active agent in a three-dimensional flow may use out-of-plane motion to move around or along such surfaces, rather than crossing transport barriers in the same manner as in the present two-dimensional setting. Therefore, the LCS-based barrier-crossing picture should be interpreted as a mechanistic explanation for navigation under two-dimensional topological constraints, rather than as a direct quantitative prediction for three-dimensional flows.

Finally, the present particle model focuses on translational dynamics. Incorporating rotational degrees of freedom, orientational response times and the simultaneous control of translational and angular accelerations would be a useful extension towards more realistic models of artificial swimmers. Beyond these kinematic and dynamical extensions, another open direction concerns the thermodynamics of active navigation. Although our current objective function penalises only the agent's mechanical propulsive effort, from a stochastic-thermodynamic perspective the continuous processes of sensing the turbulent environment, processing information and executing a control policy may also incur thermodynamic costs. Recent studies have shown that autonomous localisation and self-steering are subject to dissipation--accuracy trade-offs \citep{cocconi2025dissipation}. Accounting for the full thermodynamic dissipation associated with active control in complex convective flows therefore represents an important direction for future research, linking optimal navigation strategies to the physical limits of non-equilibrium active matter.

Overall, the present study suggests that reinforcement learning can be used not only as an optimisation tool, but also as a means of extracting interpretable physical strategies from complex flow environments. By connecting navigation performance with the reorganisation of turbulent transport barriers and pathways, our results provide a framework for understanding and designing energy-efficient autonomous navigation in buoyancy-driven flows.

\begin{bmhead}[Supplementary movies.]
Supplementary movies are available at \url{https://doi.org/10.1017/jfm.2026.11802}.
\end{bmhead}

\begin{bmhead}[Funding.]
This work was supported by the National Natural Science Foundation of China (NSFC) through grant nos. 12272311, 12388101, 12125204; the Young Elite Scientists Sponsorship Program by CAST (2023QNRC001); 
the Fundamental Research Funds for the Central Universities (no. D5000260269);
and the 111 project of China (project no. B17037). 
The authors acknowledge the Beijing Beilong Super Cloud Computing Co. Ltd for providing HPC resources that have contributed to the research results reported within this paper (\url{http://www.blsc.cn/}).
\end{bmhead}

\begin{bmhead}[Declaration of interests.]
The authors report no conﬂict of interest.
\end{bmhead}

\begin{bmhead}[Author ORCIDs]
Ao Xu, https://orcid.org/0000-0003-0648-2701;	Heng-Dong Xi, https://orcid.org/0000-0002-2999-2694.
\end{bmhead}

\begin{appen}

\section{Soft actor--critic reinforcement-learning algorithm}\label{sec:appSAC}

The SAC algorithm is an off-policy actor--critic deep-reinforcement-learning algorithm formulated within the maximum-entropy RL framework. The key idea of SAC is to augment the standard RL objective with an entropy term, thereby encouraging exploration while optimising the expected return. The SAC objective can be written as

\begin{equation}
  \pi^{*}_{\theta}
  =\arg\max_{\pi_{\theta}}
  \mathbb{E}_{\tau\sim\pi_{\theta}}
  \left[
  \sum_{t=0}^{\infty}
  \gamma^t
  \left\{
  r_t(s_t,a_t,s_{t+1})
  +\alpha \mathcal{H}\left[\pi_{\theta}(\cdot|s_t)\right]
  \right\}
  \right],
\end{equation}
where \(\pi_{\theta}\) denotes the policy, \(\theta\) denotes the neural-network parameters of the policy and \(\pi^{*}_{\theta}\) denotes the optimal policy. Furthermore, \(r_t\) is the reward at time step \(t\), \(s_t\) and \(a_t\) are the state and action at time step \(t\), \(\gamma\) is the discount factor, \(\alpha\) is the temperature parameter that controls the trade-off between reward maximisation and entropy maximisation and \(\mathcal{H}[\pi_{\theta}(\cdot|s_t)]\) is the entropy of the policy at state \(s_t\).

The SAC consists of three main components:
\begin{itemize}
  \item \textbf{\textit{Q}-networks.}
  The \emph{Q}-networks are neural networks that approximate the action--value function \(Q(s,a)\). The SAC employs two \emph{Q}-networks to reduce overestimation bias. The corresponding loss functions are
  
  \begin{equation}
    L(Q_{\phi,i}) =
    \mathbb{E}_{(s_t,a_t,r_t,s_{t+1})\sim \mathcal{D}}
    \left[
    \left(
    Q_{\phi,i}(s_t,a_t) - y_t
    \right)^2
    \right],
    \label{eq:q_loss}
  \end{equation}
  where
  
  \begin{equation}
    y_t =
    r_t
    + \gamma
    \left[
    \min_{i=1,2} Q_{\phi^{\text{target}},i}(s_{t+1},a^{\prime}_{t+1})
    - \alpha \log \pi_{\theta}(a^{\prime}_{t+1}|s_{t+1})
    \right]
  \end{equation}
  is the target value. Here, \(\mathcal{D}\) is the replay buffer, \(Q_{\phi^{\text{target}},i}\) denotes the \(i\)th target \emph{Q}-network, \(\phi\) and \(\phi^{\text{target}}\) are the neural-network parameters of the \emph{Q}-networks and target \emph{Q}-networks, respectively, and \(a^{\prime}_{t+1}\sim\pi_{\theta}(\cdot|s_{t+1})\) is the action sampled from the current policy at the next state.

  \item \textbf{Target \textit{Q}-networks.}
  The target \textit{Q}-networks are used to stabilise the training of the \textit{Q}-networks. The SAC maintains target \textit{Q}-networks that are updated through the soft-update mechanism:
  
  \begin{equation}
    \phi^{\text{target}} \leftarrow \tau \phi + (1-\tau)\phi^{\text{target}},
    \label{eq:target_update}
  \end{equation}
  where \(\tau\) is a small constant that controls the update rate. This procedure reduces rapid variations in the target values and improves training stability.

  \item \textbf{Policy network.}
  The policy network represents the policy \(\pi_{\theta}\). It is trained to maximise the expected return together with the policy entropy. The corresponding loss function is
  
  \begin{equation}
    L(\pi_{\theta}) =
    \mathbb{E}_{s_t\sim \mathcal{D},\, \xi \sim \mathcal{N}(0,I)}
    \left[
    \alpha \log \pi_{\theta}(\tilde{a}_{\theta}(s_t,\xi)|s_t)
    -
    \min_{i=1,2} Q_{\phi,i}(s_t,\tilde{a}_{\theta}(s_t,\xi))
    \right],
    \label{eq:policy_loss}
  \end{equation}
  where
  
  \begin{equation}
    \tilde{a}_{\theta}(s,\xi)
    =
    \tanh\left(
    \mu_{\theta}(s)+\sigma_{\theta}(s)\odot\xi
    \right)
  \end{equation}
  is the reparametrisation used to sample bounded actions from the Gaussian policy, with \(\xi\sim\mathcal{N}(0,I)\). Here, \(\mu_{\theta}(s)\) and \(\sigma_{\theta}(s)\) are the mean and standard deviation predicted by the policy network, and \(\odot\) denotes elementwise multiplication.
\end{itemize}

Moreover, we implement the automatic entropy-adjustment mechanism in SAC to adaptively tune the temperature parameter \(\alpha\) during training. The loss function for the temperature parameter is

\begin{equation}
  L(\alpha)
  =
  \mathbb{E}_{s_t\sim\mathcal{D},\,a_t\sim\pi_{\theta}}
  \left[
  -\alpha
  \left(
  \log \pi_{\theta}(a_t|s_t)+\bar{\mathcal{H}}
  \right)
  \right],
  \label{eq:temperature_loss}
\end{equation}
where \(\bar{\mathcal{H}}\) is the target entropy.

The SAC training procedure consists of the following steps:
\begin{itemize}
  \item Collect experience by interacting with the environment using the current policy.
  \item Store the collected experience in a replay buffer.
  \item Sample a mini-batch of experiences from the replay buffer.
  \item Update the temperature parameter \(\alpha\) by minimising the loss function in \eqref{eq:temperature_loss}.
  \item Update the \textit{Q}-networks by minimising the loss functions in \eqref{eq:q_loss} using the sampled mini-batch.
  \item Update the policy network by minimising the loss function in \eqref{eq:policy_loss} using the sampled mini-batch.
  \item Update the target \textit{Q}-networks using the soft-update mechanism in \eqref{eq:target_update}.
\end{itemize}

The hyperparameters used to train the SAC agent in this work are listed in table~\ref{tab:sac_hyperparameters}. The entries for the number of hidden layers and the number of units per layer specify the architecture of all neural networks used in the SAC agent, including the \textit{Q}-networks, target \textit{Q}-networks and the policy network.

\begin{table}
\centering
\begin{tabular}{lc}
Hyperparameter & Value \\
\\
Learning rate & \(3 \times 10^{-4}\) \\
Discount factor \(\gamma\) & \(0.99\) \\
Replay-buffer size & \(10^6\) \\
Batch size & \(256\) \\
Target update rate \(\tau\) & \(0.005\) \\
Target entropy \(\bar{\mathcal{H}}\) & \(-2\) \\
Hidden layers & \(2\) \\
Hidden units per layer & \(256\) \\
Activation function & ReLU \\
\end{tabular}
  \caption{Hyperparameters used to train the SAC agent.}
\label{tab:sac_hyperparameters}
\end{table}

\section{Ablation study on the observation state space} \label{sec:appObservation}

To assess the contribution of different sensory inputs, we conducted a systematic comparison across multiple observation sets. Specifically, we considered four baseline sets:

\begin{equation}
\mathcal{O}_1=\{\bm{u}_f^*\},\quad
\mathcal{O}_2=\{\bm{u}_f^*,T_f^*\},\quad
\mathcal{O}_3=\{\bm{u}_f^*,\bm{u}_p^*\} \quad \text{and} \quad
\mathcal{O}_4=\{\bm{u}_f^*,\bm{u}_p^*,T_f^*\}.
\end{equation}
For each baseline, we then considered three augmentations: the inclusion of spatial information $\mathcal{X}=\{\bm{x}_p^*\}$, the inclusion of gradient information \(\nabla \mathcal{O}_i\) (including \(\nabla \bm{u}_f\) and \(\nabla T_f\), where applicable) and the inclusion of both. This yields 16 combinations in total. The resulting success rates \(S\), evaluated at the corresponding optimal values of \(\mathcal{A}_{\max}\) for \(Ra=10^8\) and \(10^{10}\), are shown in figure~\ref{fig:ablation_success_rate}. These results show that both positional and gradient information improve navigation performance, in agreement with the findings of \citet{Jiao2025Gradients}. The highest success rates are achieved when both are included simultaneously, indicating that gradient information is not redundant with the basic flow observations.

\begin{figure}
    \centering
    \includegraphics[width=\textwidth]{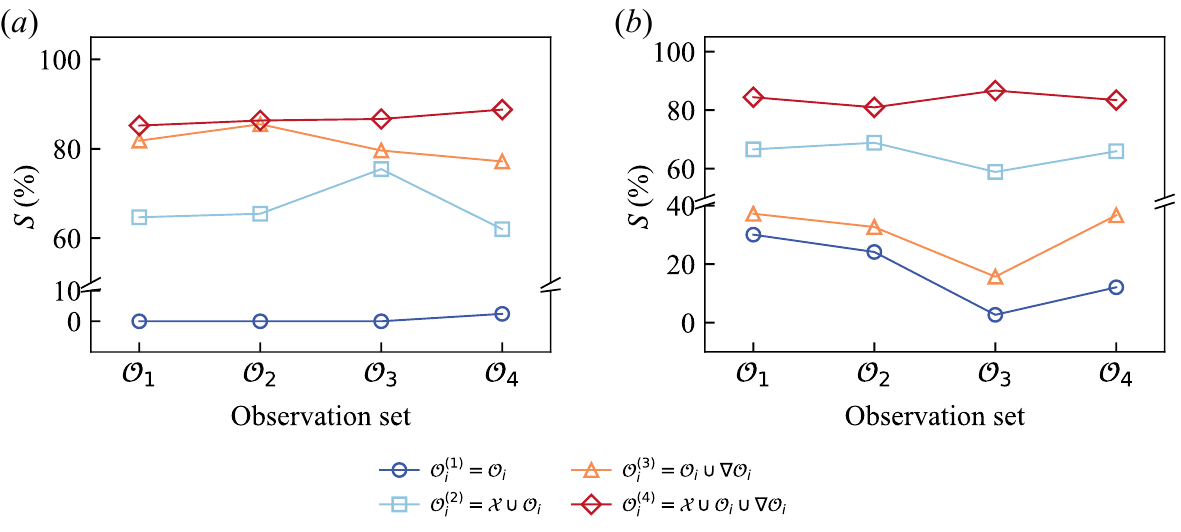}
  \caption{Success rate \(S\) for different observation state spaces at (\textit{a}) \(Ra=10^8\) and (\textit{b}) \(Ra=10^{10}\), evaluated at the corresponding optimal values of \(\mathcal{A}_{\max}\). Four baseline observation sets are considered: \(\mathcal{O}_1=\{\boldsymbol{u}_f^*\}\), \(\mathcal{O}_2=\{\boldsymbol{u}_f^*,T_f^*\}\), \(\mathcal{O}_3=\{\boldsymbol{u}_f^*,\boldsymbol{u}_p^*\}\) and \(\mathcal{O}_4=\{\boldsymbol{u}_f^*,\boldsymbol{u}_p^*,T_f^*\}\). For each baseline set \(\mathcal{O}_i\), three augmentations are considered: the inclusion of spatial information \(\mathcal{X}\), the inclusion of gradient information \(\nabla \mathcal{O}_i\) and the inclusion of both. This yields four configurations per baseline: \(\mathcal{O}_i^{(1)}=\mathcal{O}_i\), \(\mathcal{O}_i^{(2)}=\mathcal{X}\cup\mathcal{O}_i\), \(\mathcal{O}_i^{(3)}=\mathcal{O}_i\cup\nabla\mathcal{O}_i\) and \(\mathcal{O}_i^{(4)}=\mathcal{X}\cup\mathcal{O}_i\cup\nabla\mathcal{O}_i\), giving 16 configurations in total.}
    \label{fig:ablation_success_rate}
\end{figure}

Furthermore, the relative importance of these observations depends on the flow regime. At lower \(Ra=10^8\), the flow is more coherent and spatially organised, so gradient-based local cues are more strongly correlated with the large-scale dynamics, yielding a clear performance gain. At higher \(Ra=10^{10}\), however, the flow becomes more intermittent, and local gradient information alone is less sufficient for the agent to infer its global progress towards the target. In this regime, explicit positional information (\(\mathcal{X}\)) becomes more valuable for maintaining global orientation. These results suggest that both spatial coordinates and fluid gradients contribute to effective autonomous navigation in spatiotemporally chaotic environments.

To assess algorithmic stability and verify that the performance variations observed in the ablation study are not artefacts of insufficient training time, we present sample learning curves in figure~\ref{fig:learning_curves_ra1e8} (\(Ra=10^8\)) and figure~\ref{fig:learning_curves_ra1e10} (\(Ra=10^{10}\)). These curves show the evolution of the averaged cumulative reward during training. For observation sets that provide sufficient physical context, for example configurations incorporating both spatial coordinates and local fluid gradients, the reward rises rapidly and reaches a stable high-value plateau within approximately \(1\times 10^5\) training steps. By contrast, policies restricted to less informative observation sets, for example those using only local velocity, exhibit reward curves that remain at low values and fluctuate strongly, with no clear sign of convergence. Since all agents were trained for a total of \(1\times 10^6\) training steps, these curves indicate that the training duration was sufficient for the cases considered. The inability of the restricted agents to achieve high rewards therefore appears to arise from the absence of key physical information for navigating the spatiotemporally chaotic flow, rather than from premature termination of training.

\begin{figure}
    \centering
    \includegraphics[width=0.85\textwidth]{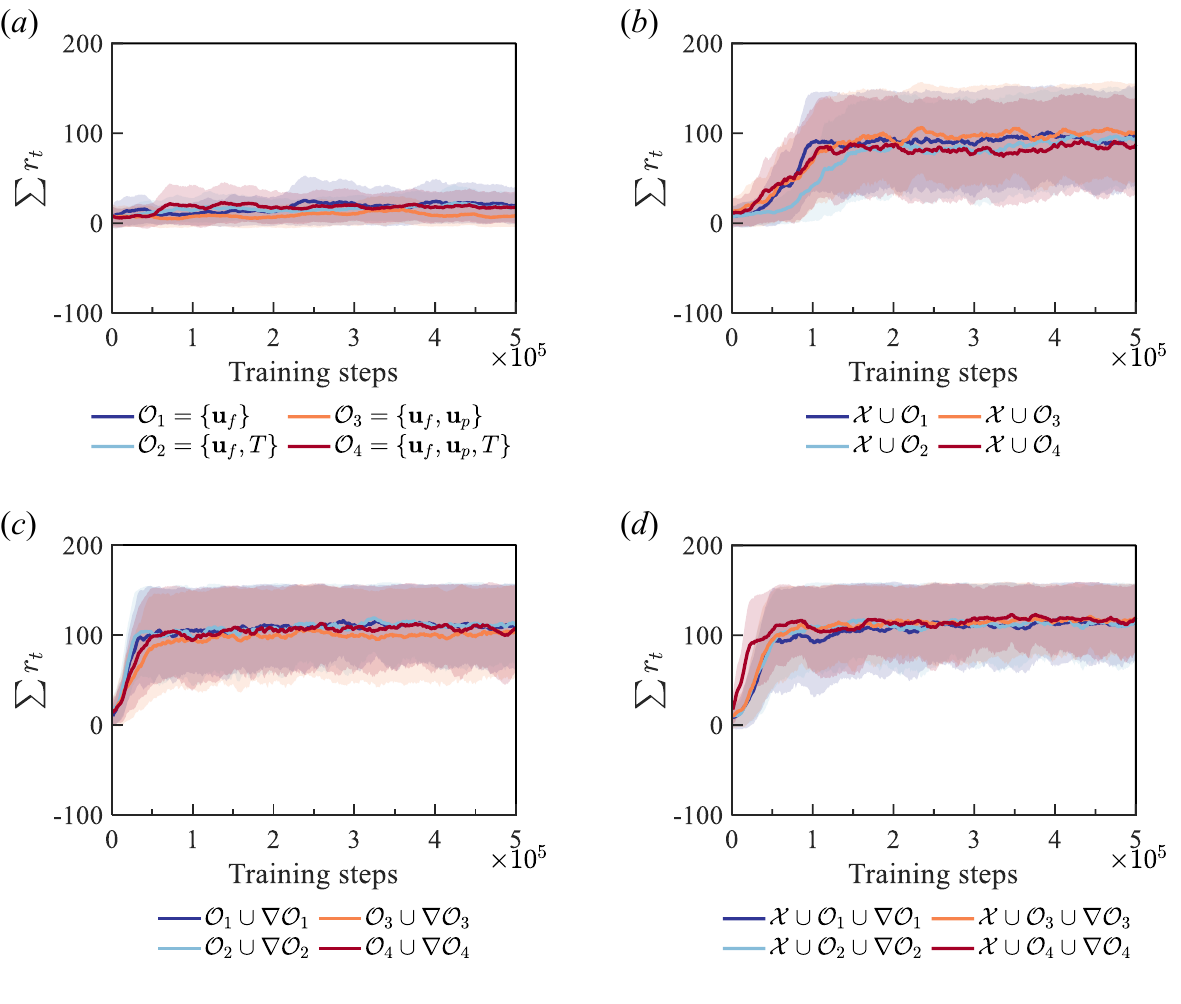}
  \caption{Sample learning curves showing the averaged cumulative reward \(\sum r_t\) as a function of training time steps for \(Ra=10^8\). The panels compare training convergence for different observation sets: (\textit{a}) the four baseline sets \(\mathcal{O}_i\), (\textit{b}) baselines augmented with spatial information \(\mathcal{X}\), (\textit{c}) baselines augmented with gradient information \(\nabla\mathcal{O}_i\) and (\textit{d}) baselines augmented with both \(\mathcal{X}\) and \(\nabla\mathcal{O}_i\). Shaded regions represent the standard deviation across training seeds.}
    \label{fig:learning_curves_ra1e8}
\end{figure}

\begin{figure}
    \centering
    \includegraphics[width=0.85\textwidth]{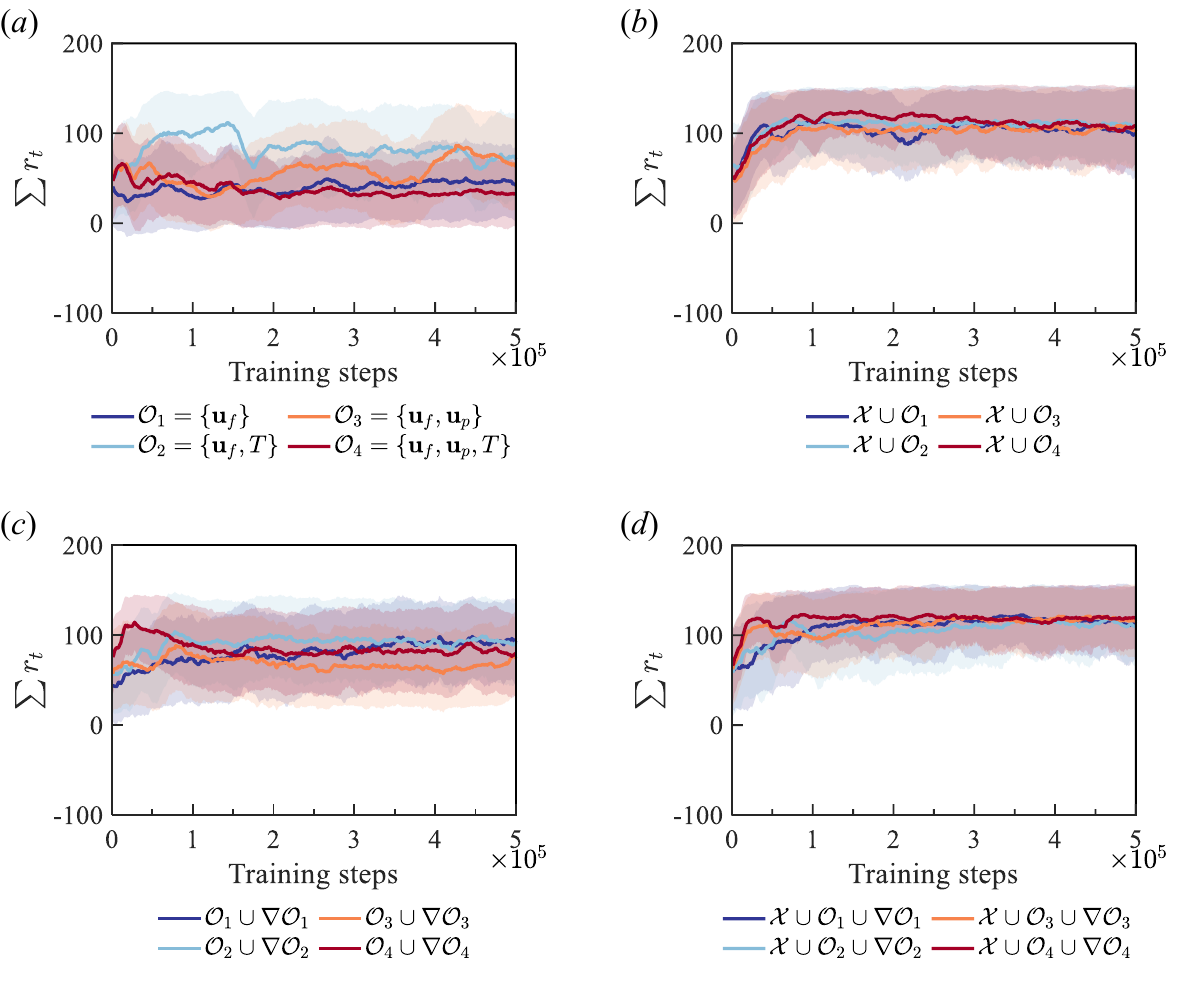}
  \caption{Sample learning curves showing the averaged cumulative reward \(\sum r_t\) as a function of training time steps for \(Ra=10^{10}\). The panels compare training convergence for different observation sets: (\textit{a}) the four baseline sets \(\mathcal{O}_i\), (\textit{b}) baselines augmented with spatial information \(\mathcal{X}\), (\textit{c}) baselines augmented with gradient information \(\nabla\mathcal{O}_i\) and (\textit{d}) baselines augmented with both \(\mathcal{X}\) and \(\nabla\mathcal{O}_i\). Shaded regions represent the standard deviation across training seeds.}
    \label{fig:learning_curves_ra1e10}
\end{figure}

\end{appen}

\bibliographystyle{jfm}



\end{document}